\documentclass[prb,nofootinbib,preprint,floatfix]{revtex4-1}
\usepackage{graphicx}
\graphicspath{{figures/}} 
\usepackage{amsmath}
\usepackage{amsfonts}
\usepackage{amssymb}
\usepackage{dcolumn}
\usepackage{braket}
\usepackage{commath}
\usepackage{epstopdf}
\usepackage{tabularx, booktabs}
\usepackage[labelsep=quad,indention=0pt]{subfig}
\usepackage{tikz}
\usetikzlibrary{calc,positioning,fit,backgrounds}

\usepackage{color}

\begin{document}

\title{Calculation of 2D electronic band structure using matrix mechanics}
\author{R.~L.~Pavelich and F.~Marsiglio}
\email{rpavelic@ualberta.ca, fm3@ualberta.ca}
\affiliation{Department of Physics, University of Alberta, Edmonton, AB, Canada T6G 2E1}

\begin{abstract}
{We extend previous work applying elementary matrix mechanics to one-dimensional periodic arrays (to generate energy bands) to two-dimensional arrays. We generate band structures for the square lattice ``2D Kronig-Penney model" (square wells), muffin-tin potential (cylindrical wells), and Gaussian wells. We then apply the method to periodic arrays of more than one atomic site in a unit cell, in particular, the case of materials with hexagonal lattices like graphene. These straightforward extensions of undergraduate-level calculations allow students to readily determine band structures of current research interest.}
\end{abstract}

\date{\today} 
\maketitle


\section{INTRODUCTION}
\label{sec:intro}

{The canonical example of a solvable one-dimensional periodic array is the Kronig-Penney model,\cite{kronig31} which yields analytically-constrained solutions. Based on the method in Ref.~[\onlinecite{marsiglio09}], where we embed some potential of interest in another confining potential with known basis states and use matrix mechanics to find the eigenvalues, the band structure for one-dimensional potentials of arbitrary shape were shown to be readily calculable in previous work.\cite{pavelich15}} In this paper, we extend the method to two-dimensional potentials, chiefly a so-called ``two-dimensional Kronig-Penney model". A specific example relevant
to current interest is the hexagonal lattice of graphene.

In typical undergraduate and graduate introductory texts in solid-state physics, the standard treatment is to introduce the nearly-free electron model followed by the tight-binding model (some canonical texts are Ref.~[\onlinecite{kittel1996}, \onlinecite{ashcroftmermin1976}, \onlinecite{singleton2001}]). These models differ in kind from problems that students are hitherto familiar with from quantum mechanics courses, and involve some suspicious suppositions. For example, in the tight-binding model the wavefunction states are
sometimes expanded in the hydrogen atomic orbitals, and the inquiring student may justly wonder how appropriate such a 
basis is for, say, sodium, when it already beings to break down for the case of helium.\cite{hutchison2012}

Therefore, we believe there is room for this matrix mechanics approach which involves only formalisms and techniques familiar to students who have taken a senior course in quantum mechanics. We review the formalism for the one-dimensional case  and then extend it to two-dimensions in section \ref{sec:formalism}. We then apply the method to the 2D Kronig-Penney model in section \ref{sec:2dkp}, followed by the cylindrical ``muffin-tin" potential in section \ref{sec:muffin} and the 2D Gaussian well in section \ref{sec:gauss}. Returning to the Kronig-Penney case of square wells, we investigate a unit cell with two wells in section \ref{sec:twocell} before concluding with the hexagonal lattice in section \ref{sec:hexagonal}. A summary is provided in section \ref{sec:conclusion}.

\section{FORMALISM}
\label{sec:formalism}

{The methodology for an infinite square well embedding potential was developed in Ref.~[\onlinecite{marsiglio09}] and was extended to embedding potentials with boundary conditions dictated by Bloch's Theorem\cite{bloch29}  in Ref.~[\onlinecite{pavelich15}]. 
We will cover only the key points here before extending the method to two dimensions.}

\subsection{One-dimensional}

{Our embedding potential is one in which the general periodicity condition}
\begin{equation}
\psi(x+a) = \psi(x)
\label{eq:generalperiodicity}
\end{equation}
is satisfied for a ``unit cell" with length $a$, which has orthonormal plane wave basis states
\begin{equation}
\psi_n^{(0)}(x) = \sqrt{\frac{1}{a}} \exp\left[ i \frac{2\pi n}{a}x \right]
\end{equation}
where $n$ is an integer: $n = ...-2, -1, 0, 1, 2, ...$. Here we have made use of a superscript $(0)$ to signify eigenstates
and eigenvalues of the embedding potential. The eigenvalues are
\begin{equation}
E_n^{(0)} = 4 \left( \frac{n^2 \pi^2 \hbar^2}{2m_0a^2} \right) = 4n^2 E_{\text{ISW}}.
\label{eq:periodicenergies}
\end{equation}
where $E_{\text{ISW}}$ is the familiar one dimensional infinite square well ground state energy, for a well with width $a$.

We can now introduce some potential $V$ of interest, and solve the matrix diagonalization problem
\begin{equation}
\sum_{m=1}^{\infty}  H_{nm} c_m = E c_n
\label{mat}
\end{equation}
where, for $H_0 = -\frac{\hbar^2}{2m_0} \frac{\dif^{\, 2}}{\dif x^2}$,
\begin{eqnarray}\nonumber
H_{nm} &=& \bra{{\psi^{(0)}_n}} (H_0 + V) \ket{{\psi^{(0)}_m}} \\
&=& \delta_{nm} E_{n}^{(0)} + H_{nm}^{V}
\label{eq:hnmv}
\end{eqnarray}
and 
\begin{eqnarray}
H_{nm}^{V} &= &{\bra{\psi^{(0)}_n} V \ket{\psi^{(0)}_m} }\nonumber \\
                     & = & {1 \over a} \int_0^a \, \dif x \, \exp{\left({-i \frac{2\pi n}{a}x}\right)} V(x) \exp{\left( i \frac{2\pi m}{a}x \right)}.
\label{eq:matrix_ele}
\end{eqnarray}
In practice we actually compute the dimensionless matrix elements $h_{nm} \equiv H_{nm}/E_{\text{ISW}}$ and the dimensionless eigenenergies $E_n^{(0)}/E_{\text{ISW}}$.

To move from a single unit cell to a periodic array, we make use of Bloch's Theorem which modifies Eq.~\ref{eq:generalperiodicity} to 
\begin{equation}
\psi(x+a) = e^{iKa} \psi(x)
\label{eq:blochstheorem}
\end{equation}
where $K$ is a wavevector satisfying the constraint $-\pi \leq Ka \leq \pi$. Here $K$ varies continuously as our potential is taken to be infinite in extent. As outlined in Ref.~[\onlinecite{pavelich15}], the effect of introducing Bloch's Theorem is to modify Eq.~\ref{eq:periodicenergies} to
\begin{eqnarray}\nonumber
E_n^{(0)} &=& \frac{\hbar^2 \pi^2}{2m_0a^2} \left( 2n + \frac{Ka}{\pi} \right)^2 \\
&=& \, \, E_{\text{ISW}} \, \, \left( 2n + \frac{Ka}{\pi} \right)^2
\label{eq:blochmodify}
\end{eqnarray}
but, remarkably, there is \textit{no} effect on Eq.~\ref{eq:matrix_ele}. That is, the imposition of Bloch's Theorem only introduces additive terms to the kinetic energy components on the main diagonal of the Hamiltonian matrix, and so the matrix need be populated only once for a given periodic potential and then repeatedly solved for different values of $Ka$ to generate the electronic band structure.

\subsection{Two-dimensional}

In two dimensions we introduce a rectangular unit cell with side lengths $a_x$ and $a_y$ obeying the general periodicity conditions
\begin{align}
\psi(x+a_x,y) &= \psi(x,y) \nonumber \\
\psi(x,y+a_y) &= \psi(x,y).
\label{eq:2dperiodicity}
\end{align}
By separation of variables and the equivalent argument as in one dimension, we have basis states
\begin{equation}
\psi_{n_xn_y}^{(0)}(x,y) = \frac{1}{\sqrt{a_xa_y}} \exp\left[ i\frac{2\pi n_x}{a_x}x + i\frac{2\pi n_y}{a_y}y\right]
\label{eq:2dbasisstates}
\end{equation}
where $n_x$ and $n_y$ are integers, with energy eigenvlaues
\begin{equation}
E_{n_xn_y}^{(0)} = 4\left[n_x^2 + n_y^2 \left( \frac{a_x^2}{a_y^2}\right) \right] E_{\text{ISW}} = E_{n_x}^{(0)} + E_{n_y}^{(0)}
\label{eq:2denergies}
\end{equation}
by analogy with Eq.~\ref{eq:periodicenergies}. The $(a_x^2/a_y^2)$ term is to account for the fact that $E_{\text{ISW}}$ is defined for length scale $a_x$. Of course, we could instead define $E_{\text{ISW}}$ in terms of $a_y$ in which case there would be an $(a_y^2/a_x^2)$ in the $E_{n_x}^{(0)}$ component. The Hamiltonian matrix elements will be of the form
\begin{align}
H_{n_xn_y,m_xm_y} &= {\bra{\psi^{(0)}_{n_xn_y}} (H_0 + V) \ket{\psi^{(0)}_{m_xm_y}} }\nonumber \\
					 &= \delta_{n_xm_x}\delta_{n_ym_y} E_{n_xn_y}^{(0)} + H_{n_xn_y,m_xm_y}^{V}
\end{align}
where $m_x$ and $m_y$ are also integers.

In order to impose the Bloch condition, we modify Eq.~\ref{eq:2dperiodicity} to
\begin{align}
\psi(x+a_x,y) &= e^{iK_x a_x}\psi(x,y) \nonumber \\
\psi(x,y+a_y) &= e^{iK_y a_y}\psi(x,y)
\label{eq:2dperiodicitybloch}
\end{align}
as was discussed in Section IV.C of Ref.~[\onlinecite{pavelich15}]. Like the one-dimensional problem, the Bloch condition will only affect the kinetic energy terms. Following Eq.~\ref{eq:blochmodify}, we modify Eq.~\ref{eq:2denergies} to
\begin{align}
E_{n_x}^{(0)} =& E_{\text{ISW}} \left( 2n_x + \frac{K_xa_x}{\pi} \right)^2 \nonumber \\
E_{n_y}^{(0)} =& E_{\text{ISW}} \left( 2n_y + \frac{K_ya_y}{\pi} \right)^2 \left( \frac{a_x^2}{a_y^2}\right).
\label{eq:2denergyscaled}
\end{align}
We have separated the two energy components here (and in later sections) for clarity, but of course there is only one summed energy $E_{n_xn_y}^{(0)}$, as in Eq.~\ref{eq:2denergies}. The procedure then will be to compute the $K_x = K_y = 0$ Hamiltonian matrix case once, and then repeatedly diagonalize for different values of $K_x$ and $K_y$ which modify the Hamiltonian matrix per Eq.~\ref{eq:2denergyscaled}.

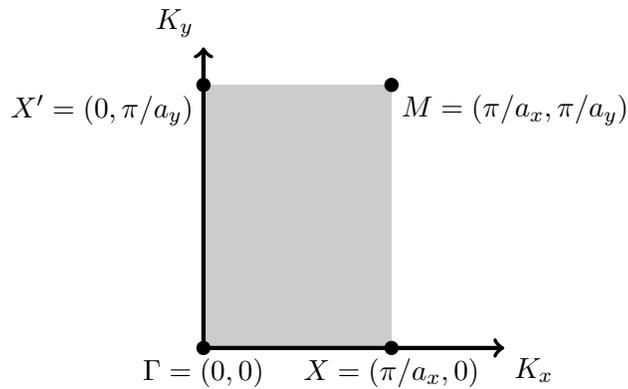
\begin{figure} [h]
\begin{tikzpicture}
	\fill [black!20] (0,0) rectangle (2.5, 3.5);

	\draw [ultra thick, <->] (0,4) -- (0,0) -- (4,0);
	\node [below right] at (4,0) {$K_x$};
	\node [above left] at (0,4) {$K_y$};
	\draw [fill] (0, 0) circle [radius=2.5pt];
	\node [below ] (0, 0) {$\Gamma = (0,0)$};
	\draw [fill] (2.5, 0) circle [radius=2.5pt];
	\node [below] at (2.5, 0) {$X = (\pi/a_x,0)$};
	\draw [fill] (0, 3.5) circle [radius=2.5pt];
	\node [below left] at (0, 3.5) {$X' = (0, \pi/a_y)$};
	\draw [fill] (2.5, 3.5) circle [radius=2.5pt];
	\node [below right] at (2.5, 3.5) {$M = (\pi/a_x, \pi/a_y)$};
	
\end{tikzpicture}
\caption{Representation in $K$-space of the location of the high-symmetry points. Similar symmetric points are found in the second, third, and fourth quadrants (not shown).} 
\label{fig:kspace}
\end{figure}

While we can and will generate band structures for the whole ``area" of $K$-space, it is useful (and for three-dimensional lattices, necessary), to trace a one-dimensional path through the two-dimensional $K$-space hitting ``high-symmetry points" as we go. Following typical convention, we define some of these points $(K_x, K_y)$ to be $\Gamma = (0,0)$, $X = (\pi/a_x,0)$, $X' = (0,\pi/a_y)$, and $M = (\pi/a_x,\pi/a_y)$ (see Fig.~\ref{fig:kspace}). Most of the figures in this work will trace the triangular path $\Gamma \rightarrow X' \rightarrow M \rightarrow \Gamma$.

\section{2D Kronig Penney model}
\label{sec:2dkp}

{We now introduce the so-called ``2D Kronig Penney model," a straightforward extension of the one-dimensional case. In a square unit cell $a_x = a_y \equiv a$ extending from $0$ to $a$ along both axes, we introduce a well with height $V_0$ (with dimensionless value $v_0 \equiv V_0/E_{\text{ISW}}$), which will typically have a negative value, in the region}
\begin{equation}
0 \leq q_1 \leq q_2 \leq a
\end{equation}
for some fractional distances $q_1$ and $q_2$. In dimensionless form, these lengths will be normalized by the factor $a$ and so we introduce $p_1 = q_1/a$ and $p_2 = q_2/a$ such that our now dimensionless distances obey
\begin{equation}
0 \leq p_1 \leq p_2 \leq 1.
\end{equation}

The matrix elements for this potential are of the form
\begin{equation}
H_{n_xn_y,m_xm_y}^{V} = \frac{V_0}{a^2} \int_{q_1}^{q_2} \int_{q_1}^{q_2} \dif x \dif y \, e^{i2\pi \left(m_x - n_x\right)x/a} e^{i2\pi \left(m_y - n_y\right)y/a}
\end{equation} 
or in dimensionless form
\begin{equation}
h_{n_xn_y,m_xm_y}^{V} = {v_0} \int_{p_1}^{p_2} \int_{p_1}^{p_2} \dif x \dif y \, e^{i2\pi \left(m_x - n_x\right)x} e^{i2\pi \left(m_y - n_y\right)y}
\end{equation}
after making the transformation $x' = x/a$ and $y' = y/a$ and then dropping the dummy index primes for convenience. This double integration factors into a product of two one-dimensional integrals which yields
\begin{equation}
h_{n_xn_y,m_xm_y}^V = v_0 I(n_x, m_x) I(n_y, m_y)
\label{eq:2dkronigfactored}
\end{equation}
where
\begin{equation}
I(n,m) = \left( p_2 - p_1 \right) \delta_{nm} + i \frac{\left( e^{i2\pi (m-n)p_1} - e^{i2\pi (m-n)p_2}\right)}{2\pi (m-n)} \left(1 - \delta_{nm} \right).
\label{eq:2dkronigI}
\end{equation}
Notice that for the main diagonal elements, the non-kinetic contribution will simply be the volume of the well, $v_0(p_2-p_1)^2$. In general, we could position the well anywhere within the unit cell, so the above integral could also be considered a function of $p_1$ and $p_2$. In this paper we will typically use $p_1 = 1/4$ and $p_2 = 3/4$.

\begin{figure}[h]
\centering
\subfloat[$v_0 = 0$]{\includegraphics[width=0.45\textwidth]{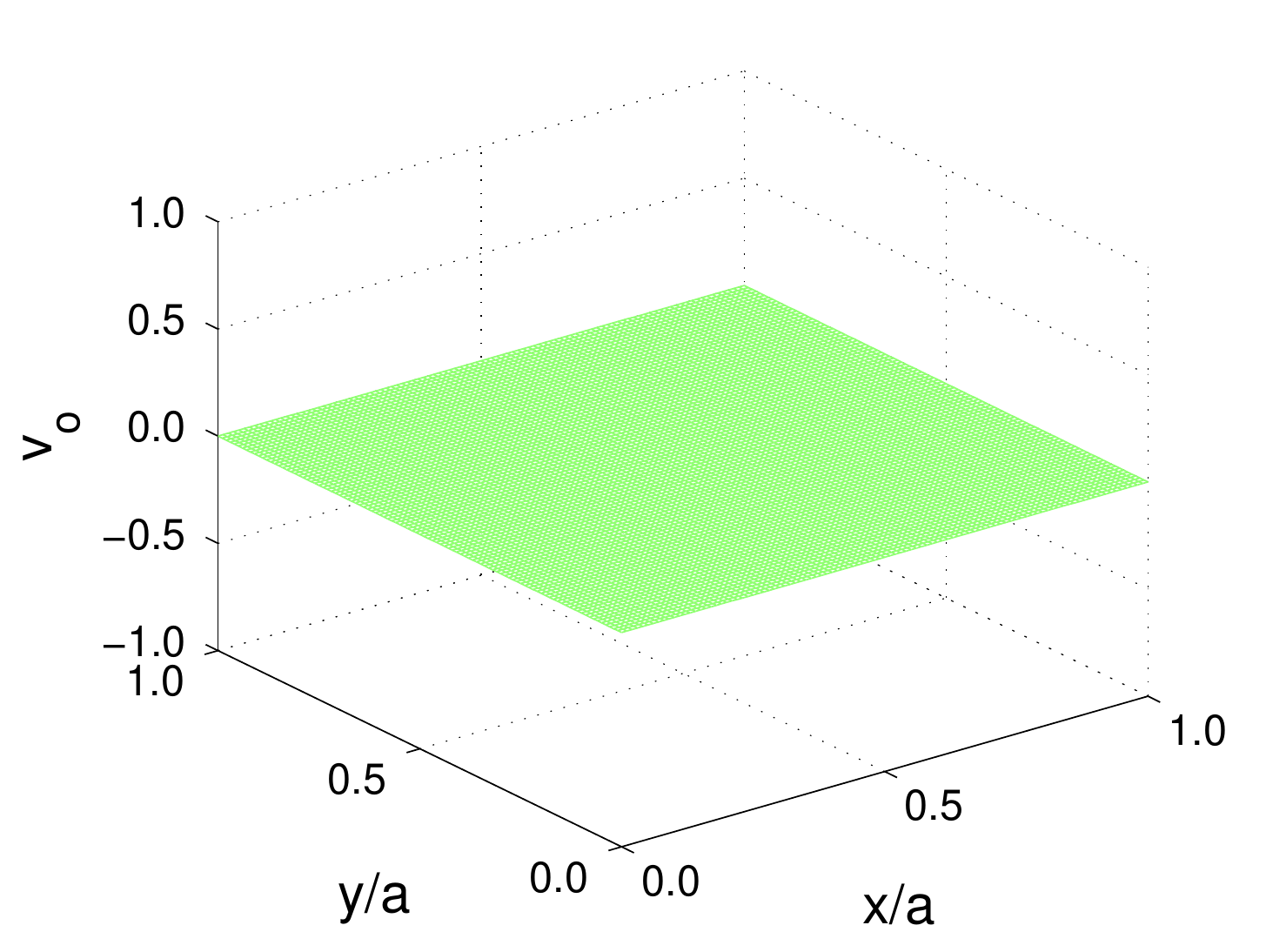}} 
\subfloat[$v_0 = 0$]{\includegraphics[width=0.45\textwidth]{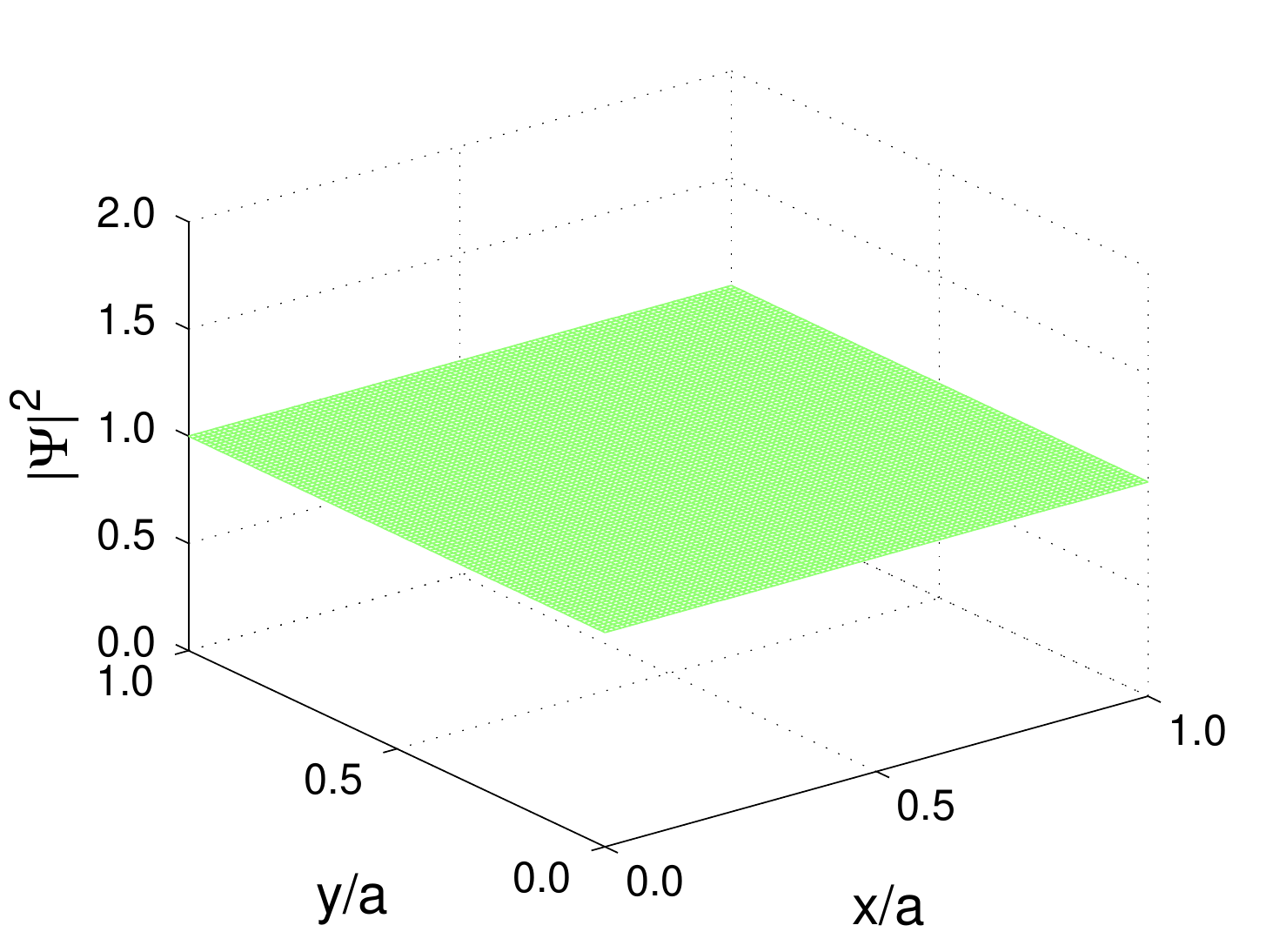}}\\
\subfloat[$v_0 = -10$]{\includegraphics[width=0.45\textwidth]{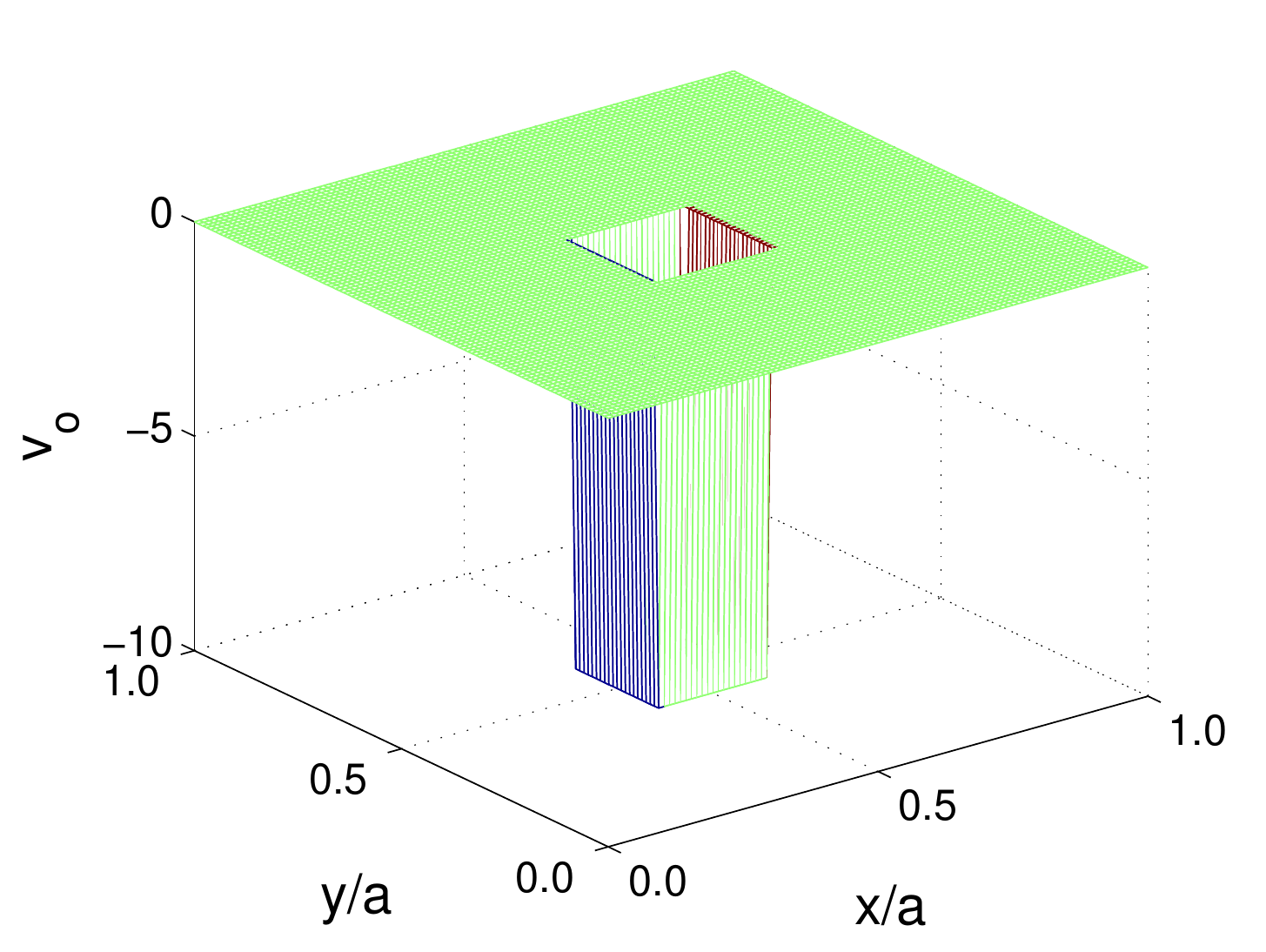}} 
\subfloat[$v_0 = -10$]{\includegraphics[width=0.45\textwidth]{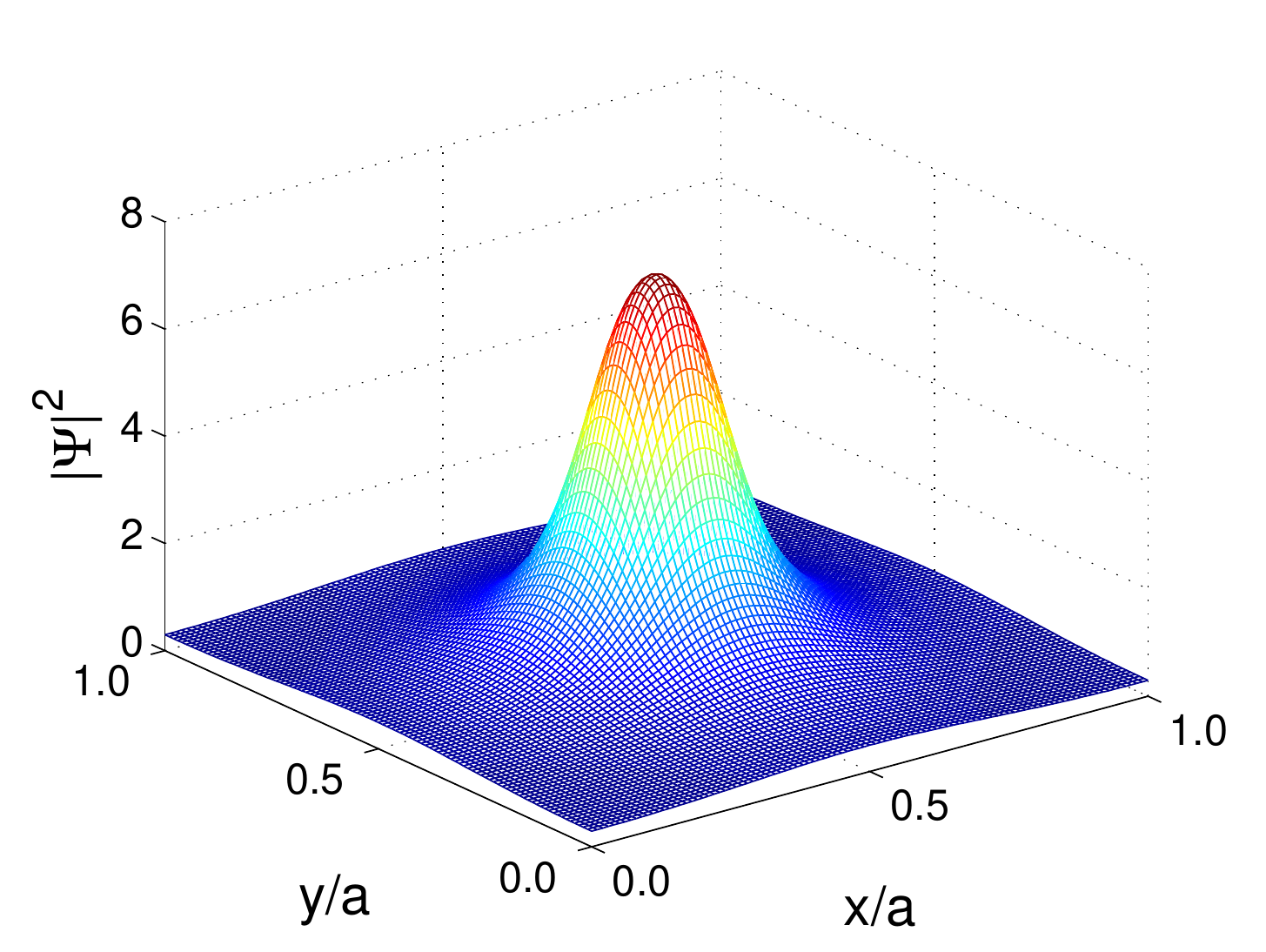}}\\
\subfloat[$v_0 = -100$]{\includegraphics[width=0.45\textwidth]{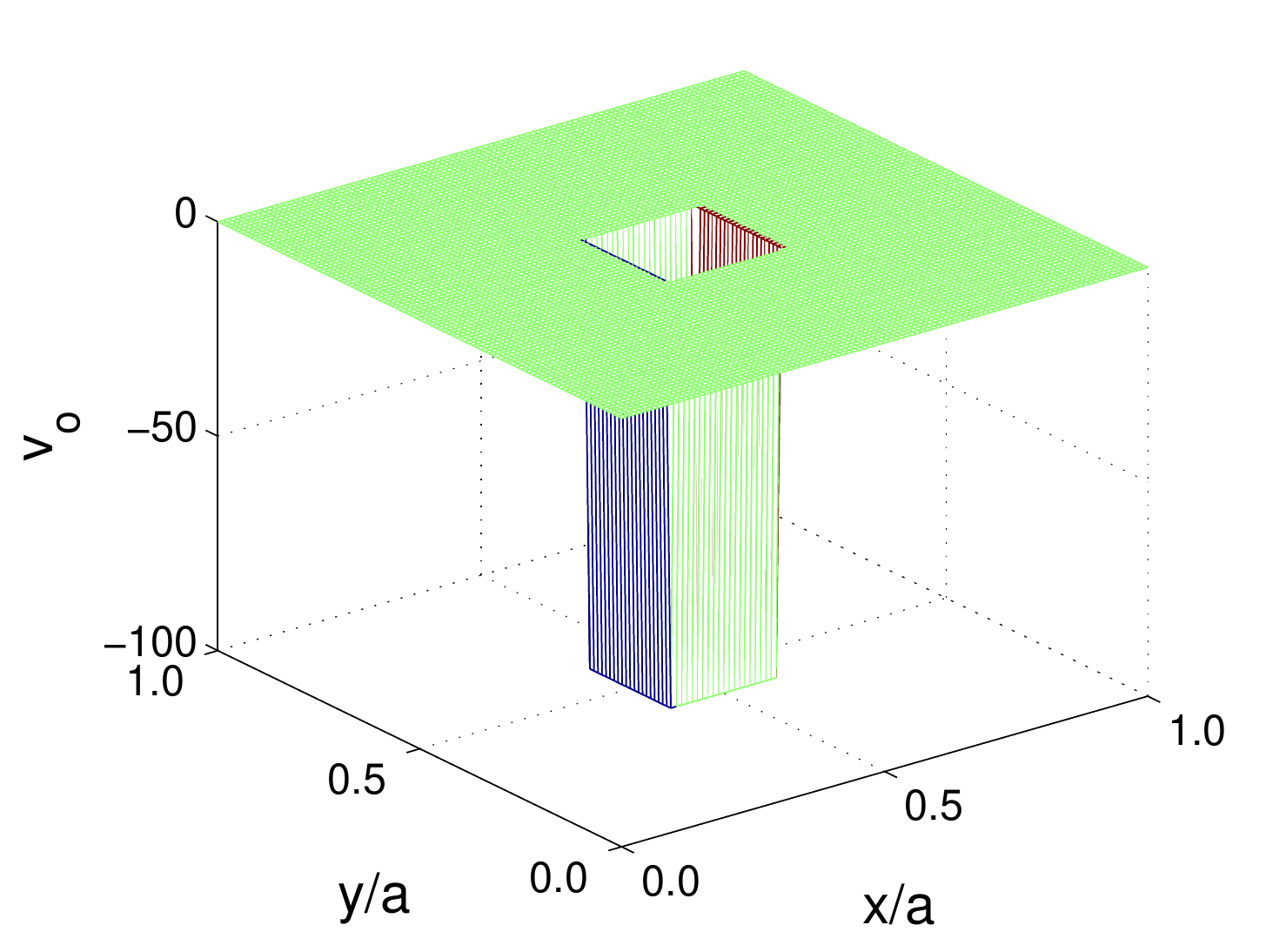}}
\subfloat[$v_0 = -100$]{\includegraphics[width=0.45\textwidth]{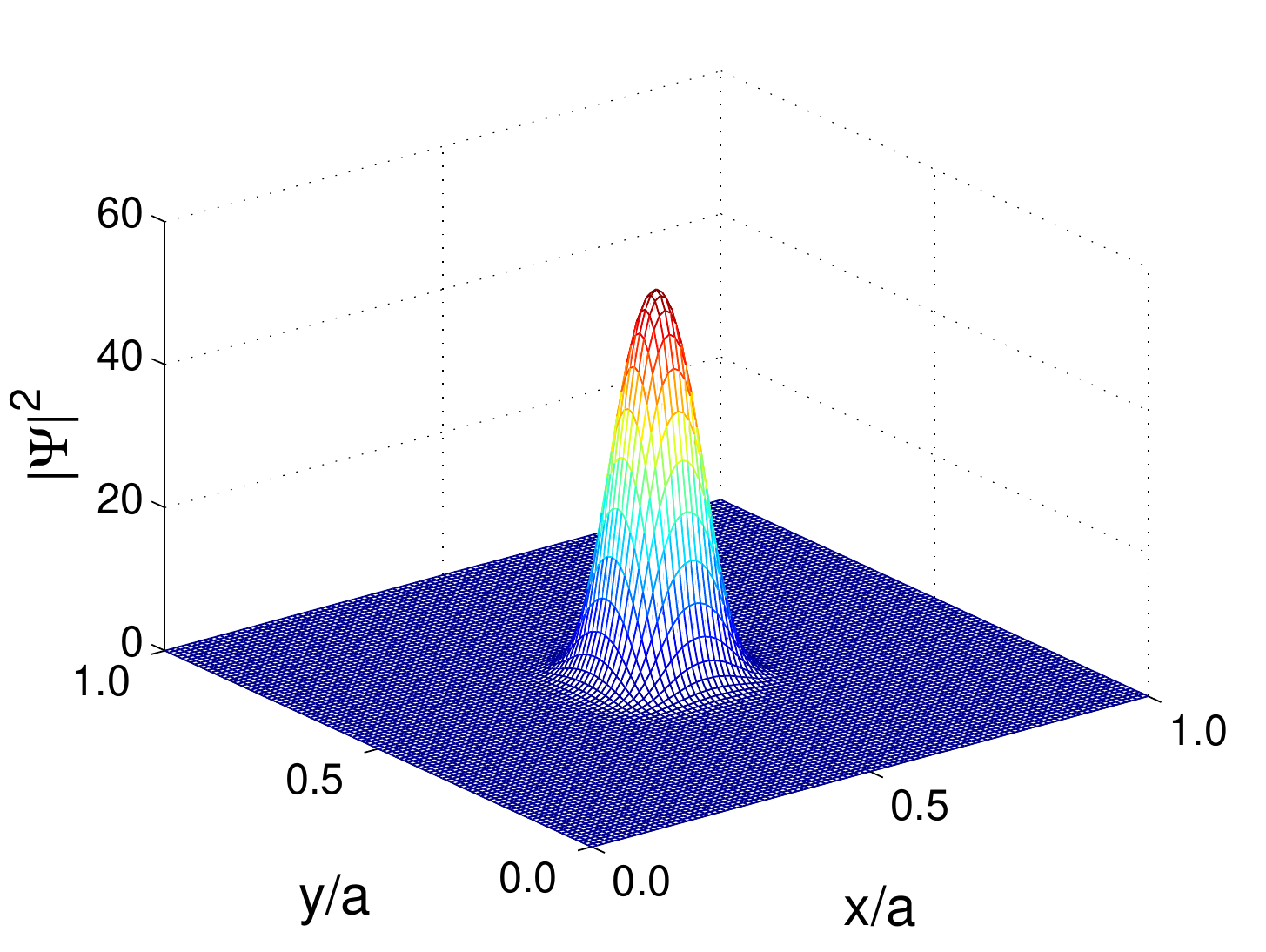}}
\caption{Representation of some 2D Kronig-Penney unit cell potentials (left column) and associated (non-Bloch modulated) wavefunctions (right column). Here $p_1 = 0.4$ and $p_2 = 0.6$.}
\label{fig:kpwaves}
\end{figure}

As a check, we used the eigenstates produced by the diagonalization for $K_x = K_y = 0$ to produce the ground state wavefunction for various well depths, shown in Fig.~\ref{fig:kpwaves}. As expected, deeper wells more tightly constrain the wavefunction.

\subsection{Nearly-free electron limit}

\begin{figure}[h]
\centering
\subfloat[$v_0 = 0$]{\includegraphics[width=0.45\textwidth]{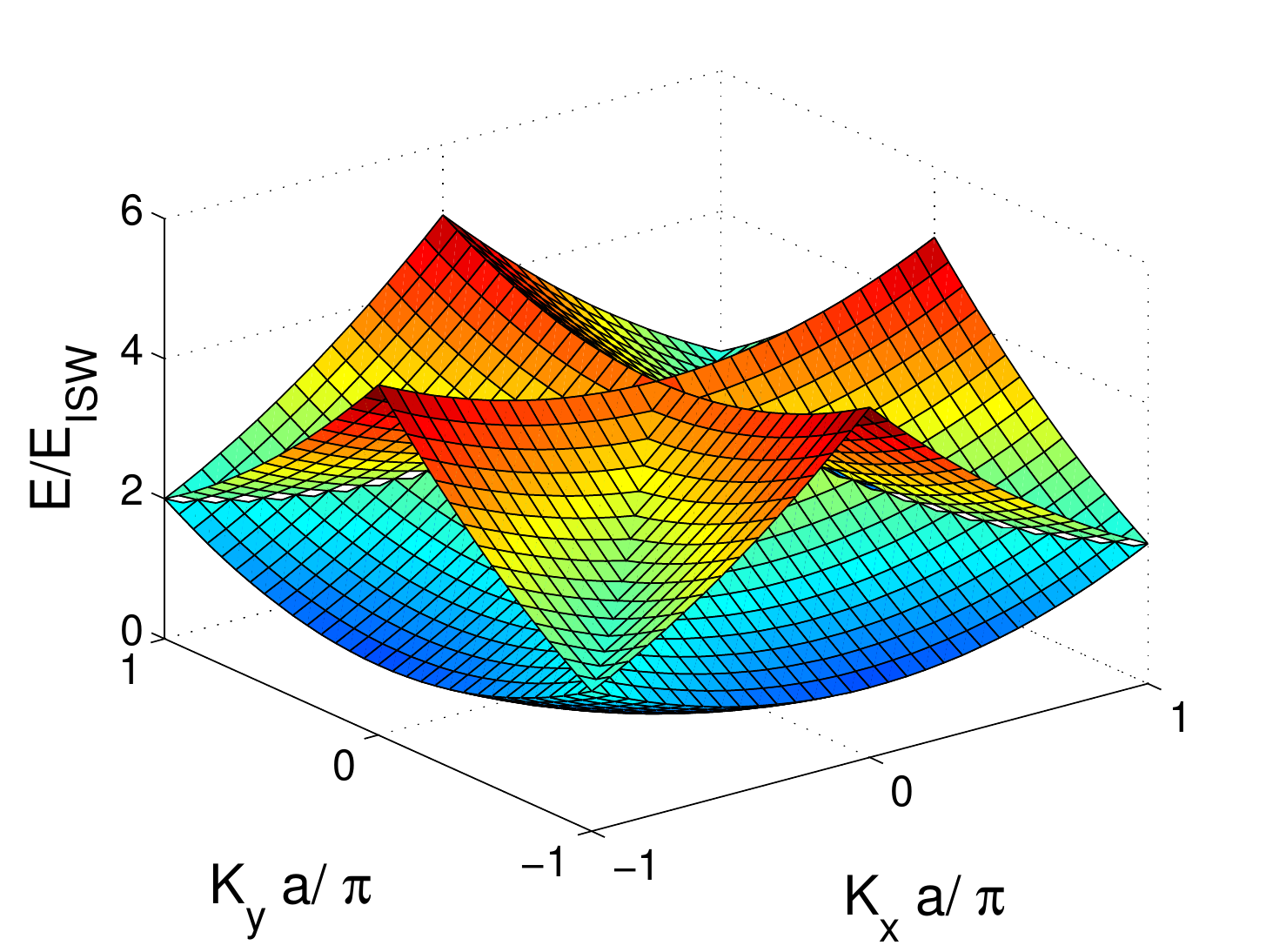}} 
\subfloat[$v_0 = 0$]{\includegraphics[width=0.45\textwidth]{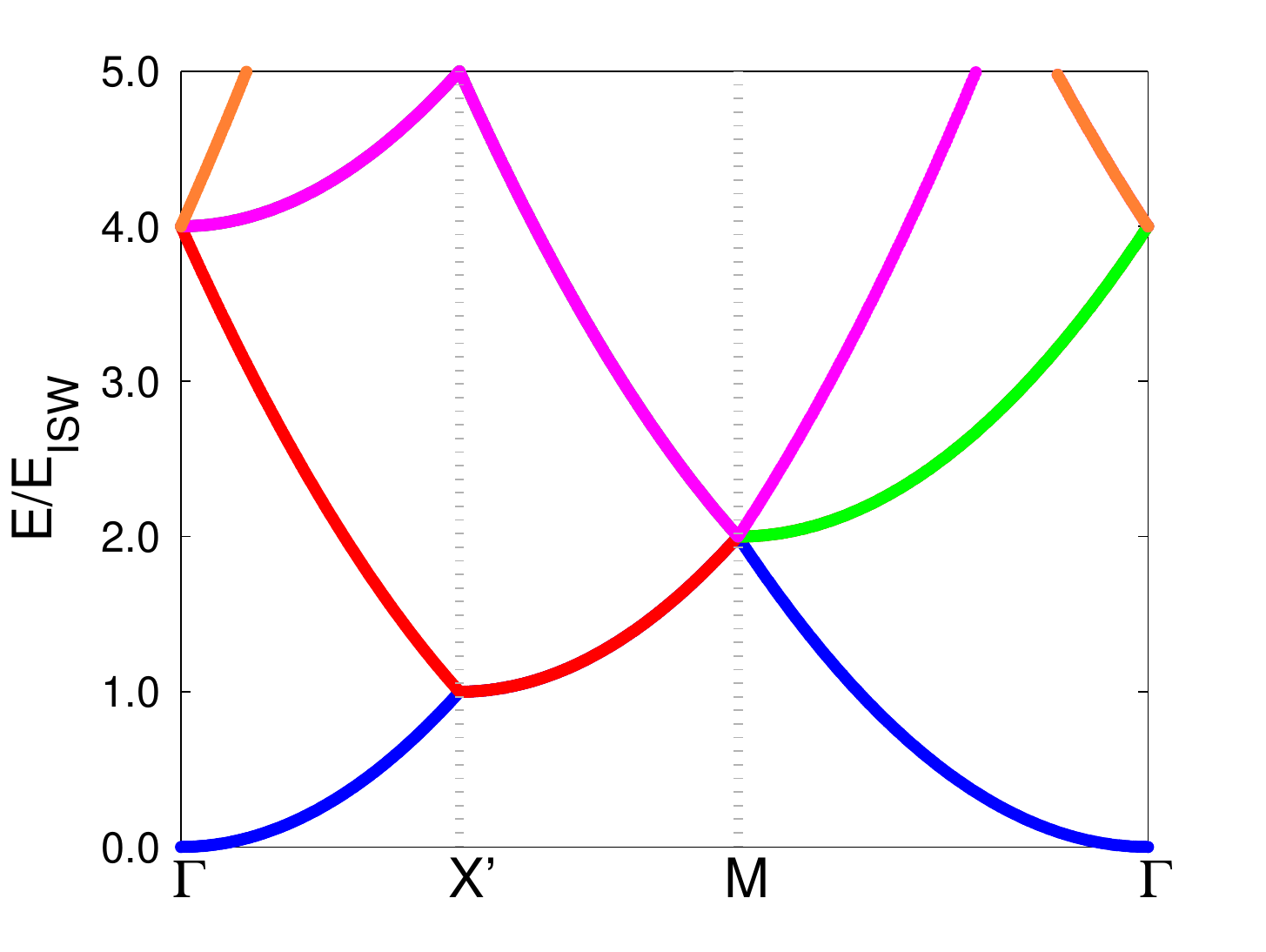}}\\
\subfloat[$v_0 = -1$]{\includegraphics[width=0.45\textwidth]{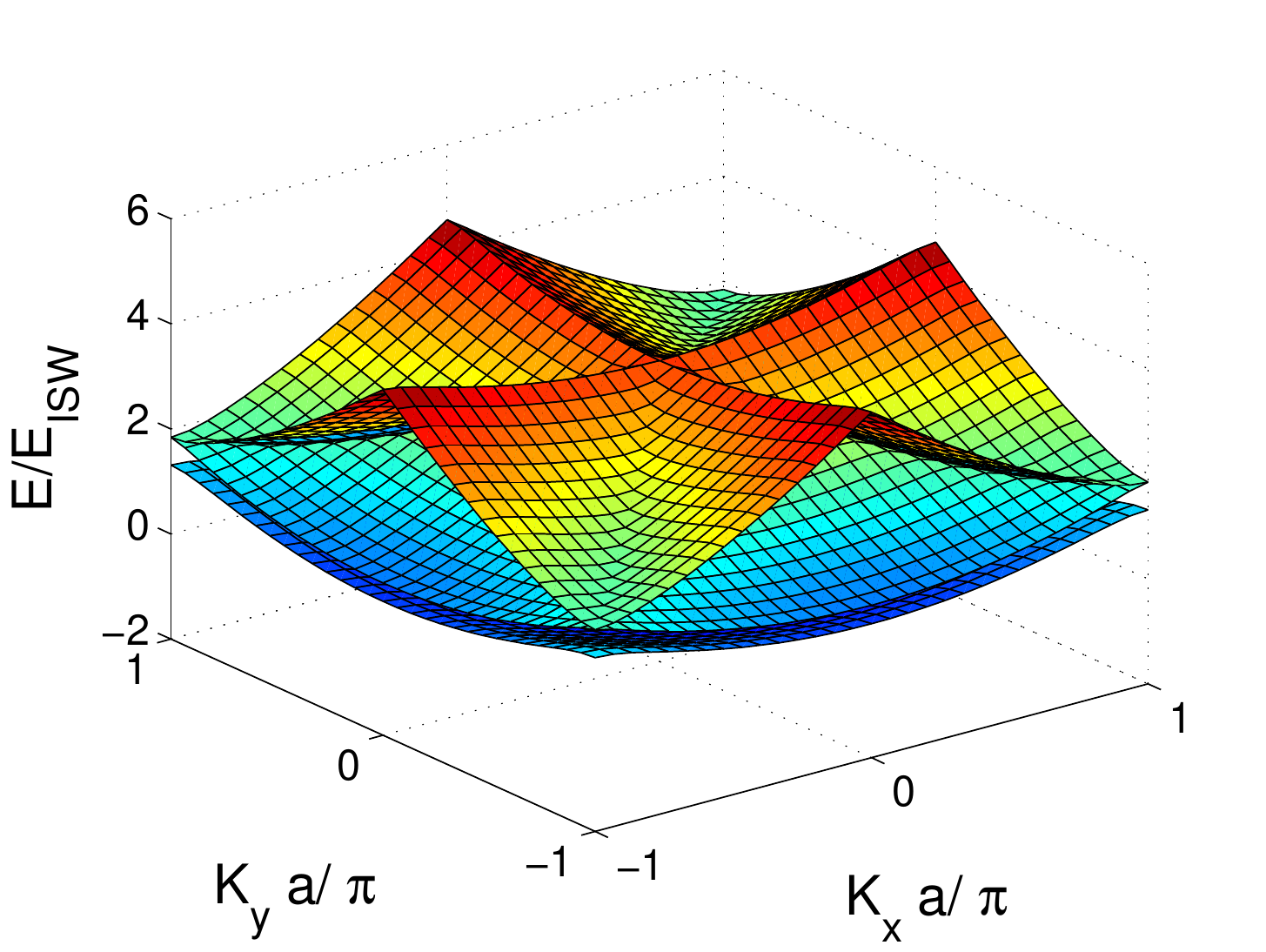}} 
\subfloat[$v_0 = -1$]{\includegraphics[width=0.45\textwidth]{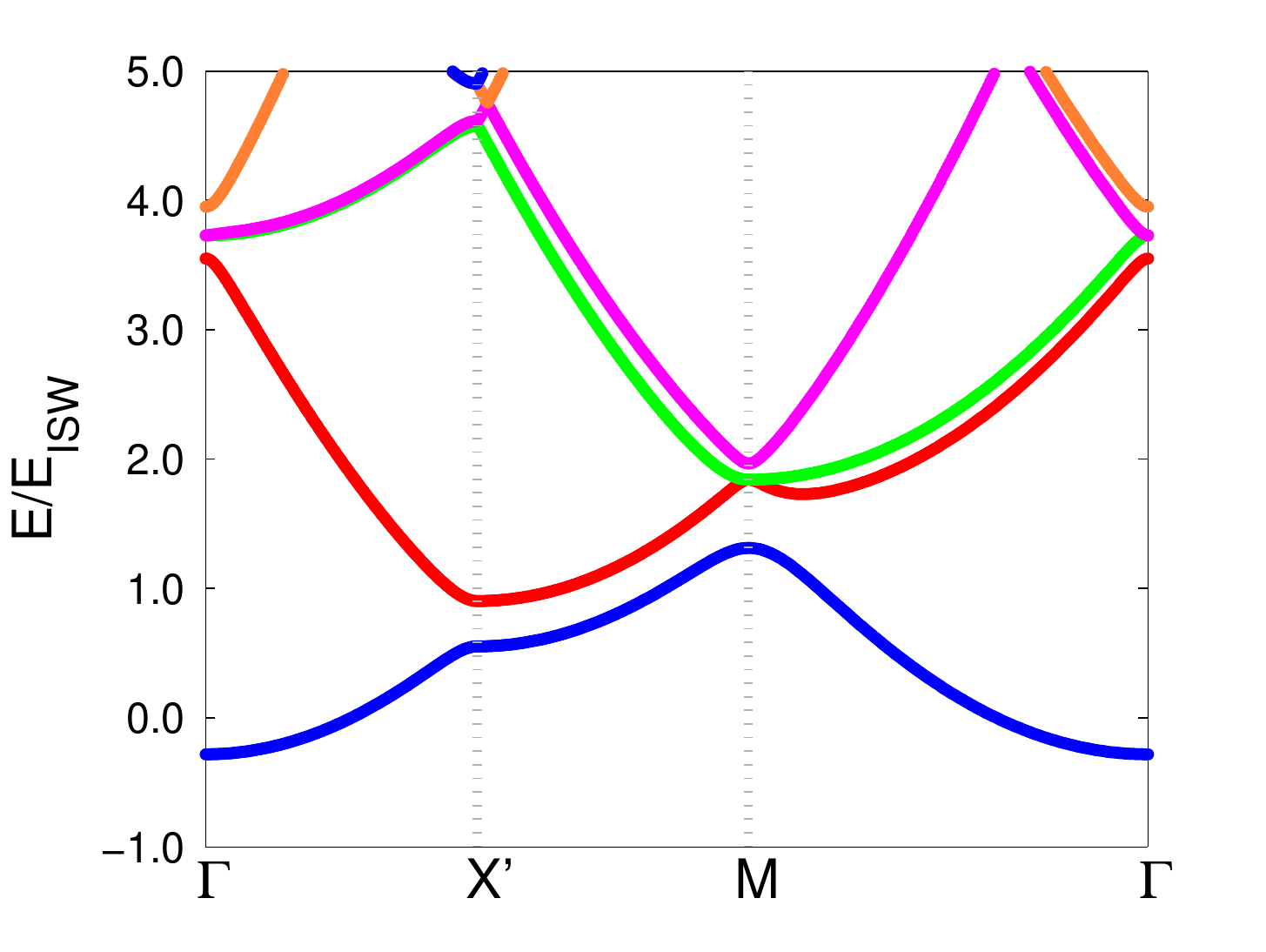}}\\
\subfloat[$v_0 = -3$]{\includegraphics[width=0.45\textwidth]{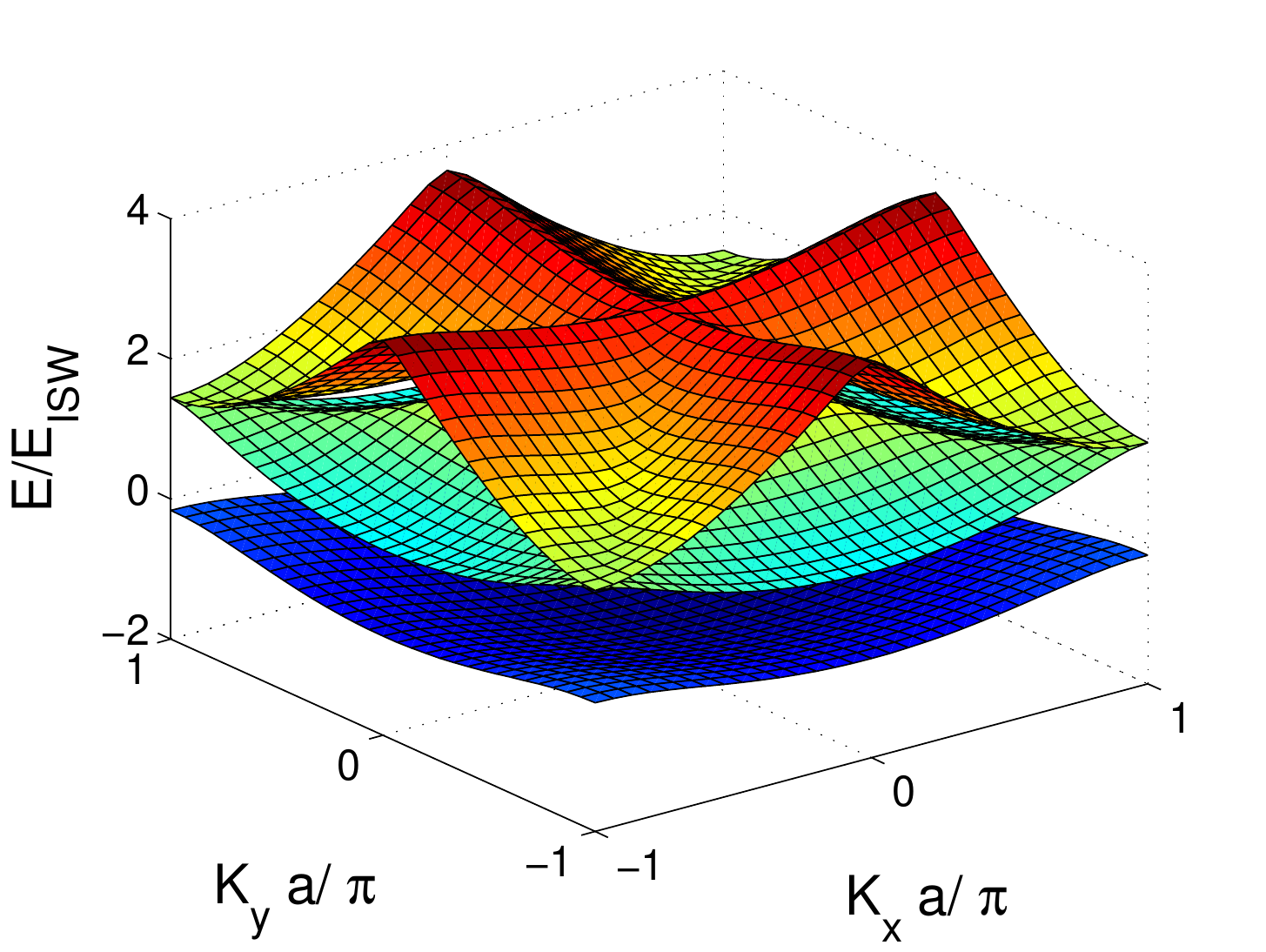}}
\subfloat[$v_0 = -3$]{\includegraphics[width=0.45\textwidth]{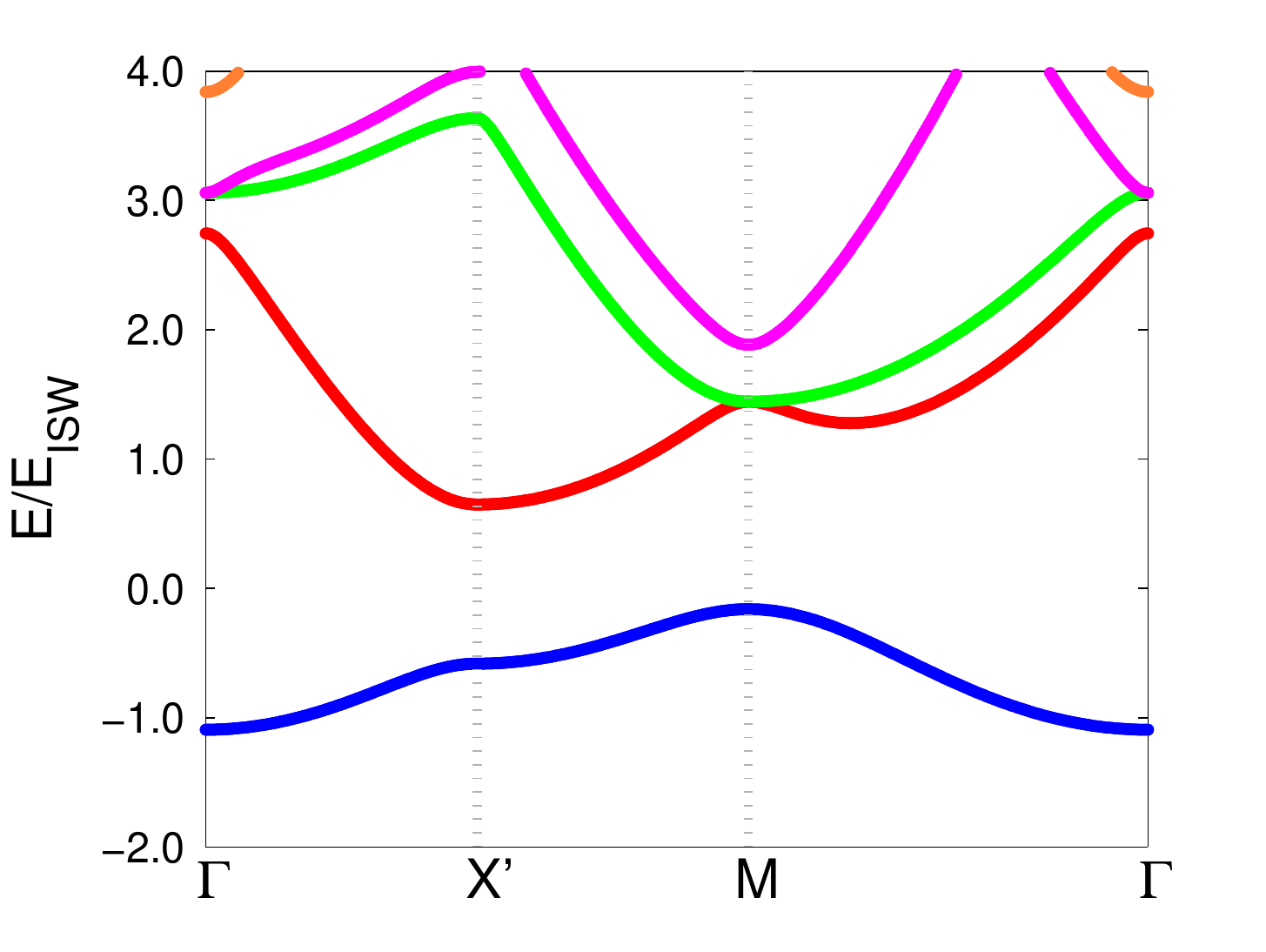}}
\caption{Generated band structures for various 2D Kronig-Penney well depths with $p_1 = 0.25$ and $p_2 = 0.75$. The full 3D representation is shown in the left column (for the first three energy bands) and the corresponding flattened plots for high-symmetry points is shown in the right column.}
\label{fig:kpbands}
\end{figure}

For $v_0 = 0$, this approach will recapitulate the so-called empty lattice approximation with parabolic free electron bands, as shown in 
Figs.~\ref{fig:kpbands}a~and~\ref{fig:kpbands}b. Further, this method provides a convenient encoding schema for the energy bands in terms of our chosen basis states. Consider the lowest energy branch going from $\Gamma$ to $X'$, ie.~along the $K_y$-axis from $0$ to  $\pi/a$ in Fig.~\ref{fig:kpbands}b, where $n_x = 0$. Referring to Eq.~\ref{eq:2denergyscaled}, we can see that this lowest branch corresponds to $n_y = 0$, and only the $K_ya$ contributes. The next higher energy branch corresponds to $n_y = -1$ as can be seen by direct substitution into the equation. Next is $n_y = +1$, and so on.

When we turn on $v_0$ for some small value, we enter the regime of the nearly-free electron model, whose salient feature is a lifting of degeneracies where band gaps emerge, and the sharp cusps for the $v_0 = 0$ case become smooth parabolas, as we can see in Fig.~\ref{fig:kpbands}d (and to a lesser extent in  Fig.~\ref{fig:kpbands}c). Finally, we show results in 
Figs.~\ref{fig:kpbands}e~and~\ref{fig:kpbands}f for an even deeper well, where an energy gap between the lowest band and the other
bands exists for all wave vectors, reminiscent of the case in one dimension.

\subsection{Tight binding model limit}

{In the opposite limit for very deep wells we consider the tight-binding model. For two-dimensional systems with horizontal and vertical symmetry and nearest-neighbor hopping the energy bands have the well-known form}
\begin{equation}
E(K_x, K_y) = -2t \left[ \cos(K_xa) + \cos(K_ya) \right]
\end{equation}
where $t$ is the so-called hopping integral. What we are interested in is not $t$ itself but the cosine behavior of the energy bands. The
question is, as we make the well deeper and deeper, does the energy band approach such a limit?

\begin{figure}[h]
\centering
\subfloat[$v_0 = 0, R^2 = 0.9234$]{\includegraphics[width=0.45\textwidth]{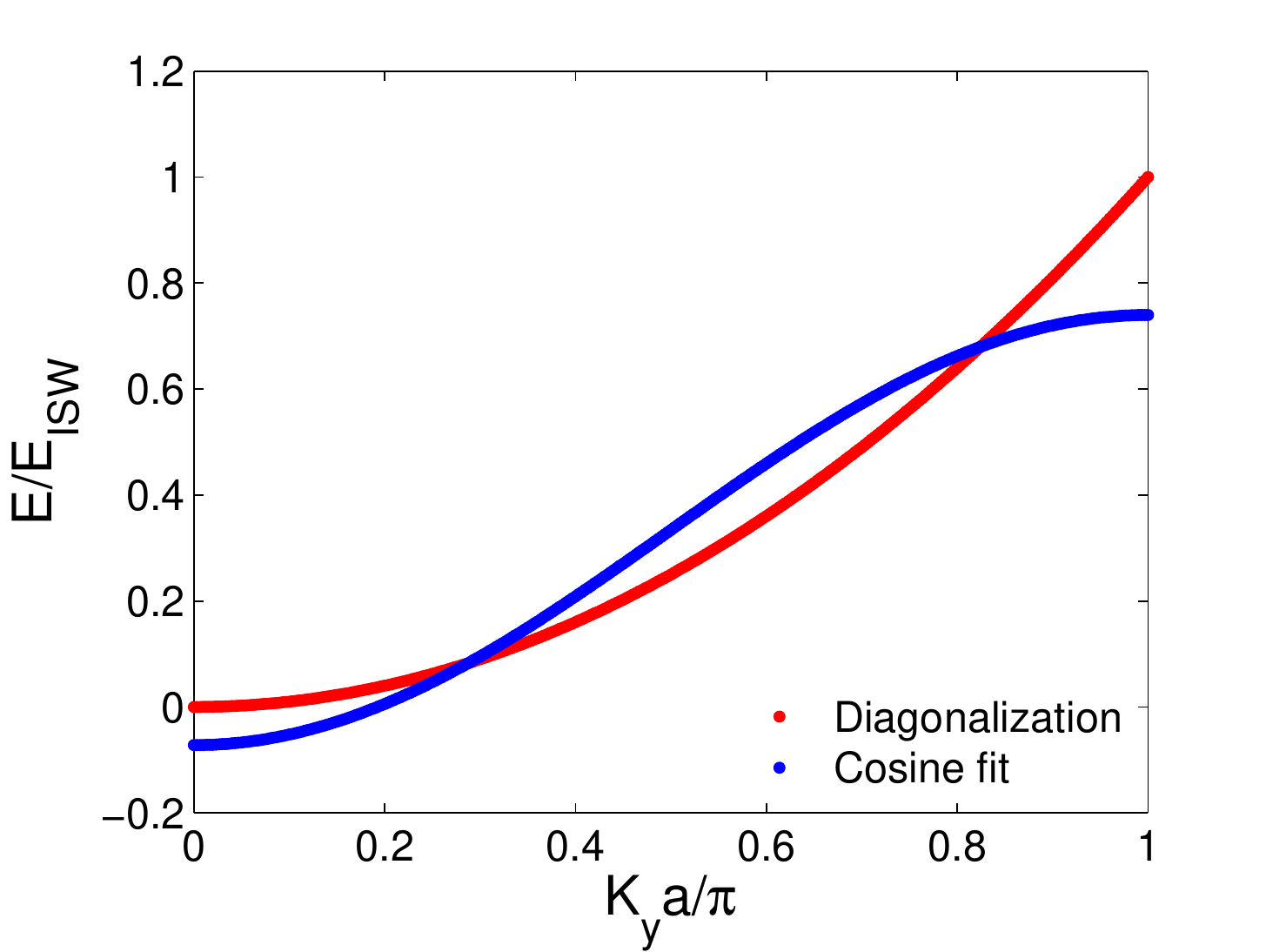}}\\
\subfloat[$v_0 = -3, R^2 = 0.9854$]{\includegraphics[width=0.45\textwidth]{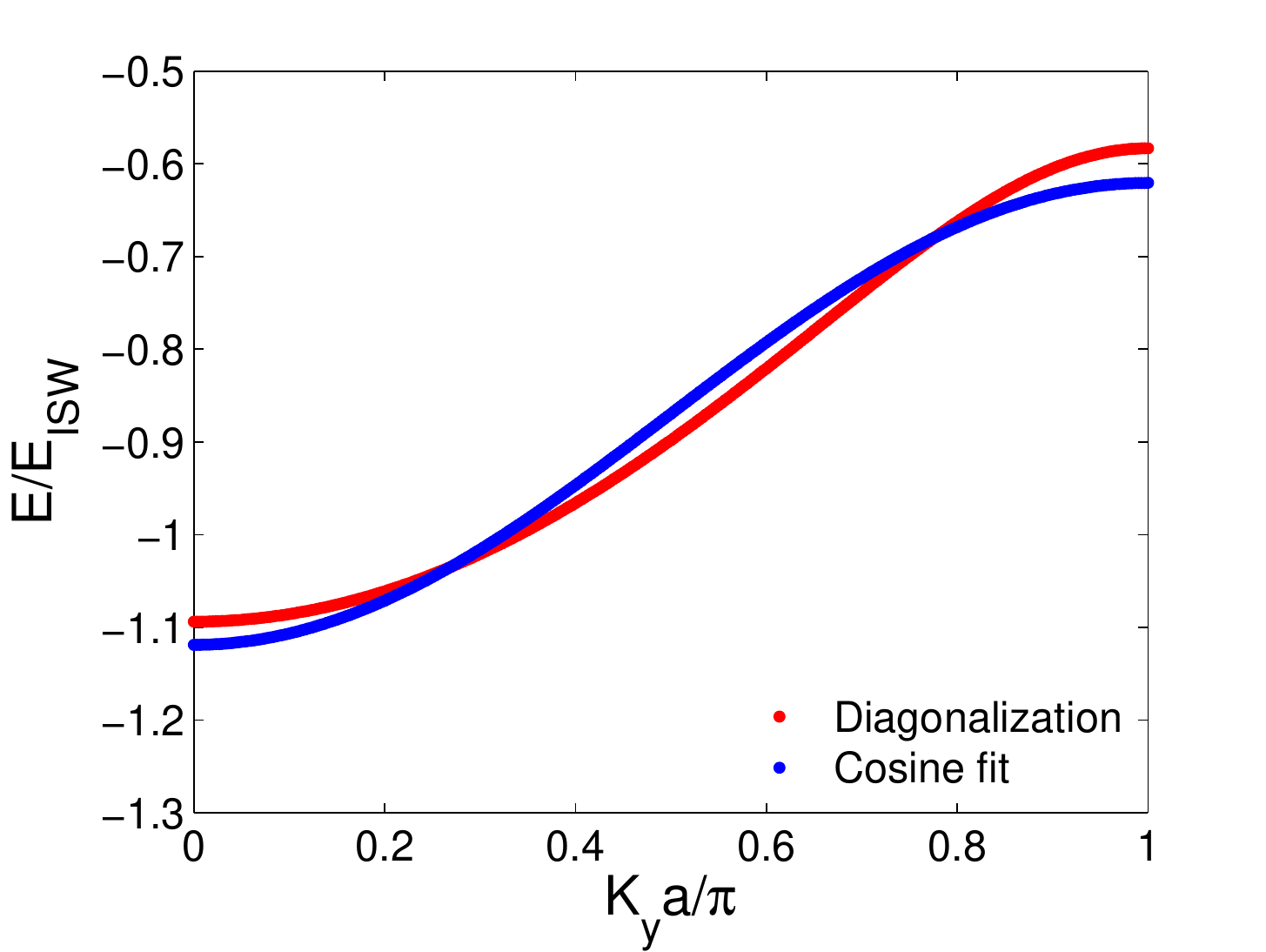}}\\
\subfloat[$v_0 = -10, R^2 = 0.9999$]{\includegraphics[width=0.45\textwidth]{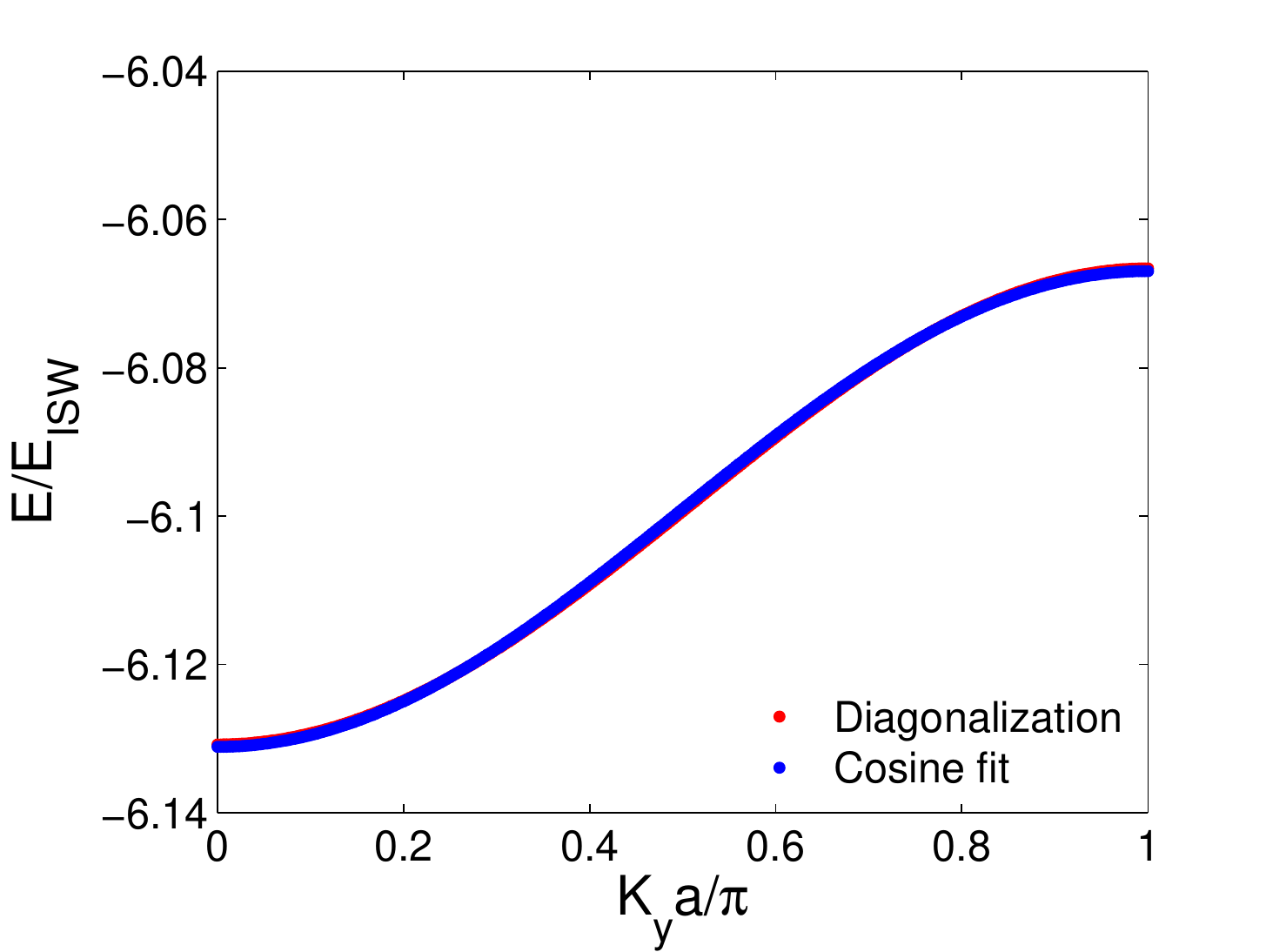}}
\caption{2D Kronig-Penney results for the lowest-energy band with a fitted tight-binding cosine function, with $p_1 = 0.25$, $p_2 = 0.75$. As the well depth $v_0$ deepens, the results approach the tight-binding limit. Note especially in (b) the asymmetry between the top and bottom of the band. In (c) the curves are essentially on top of one another.}
\label{fig:tightbind}
\end{figure}

As our potential is symmetric, it does not matter which direction in $K$-space we choose, and so we arbitrarily select the $K_y$ axis such that $K_xa = 0$. Then we numerically fit to the lowest-energy band a function $-2t\cos(\pi y) + E$ where ``$y$" is our $K_ya/\pi$ result and $E$ is a factor needed so the fitted curve is correctly situated vertically. The results are shown in Fig.~\ref{fig:tightbind}. We see that as the well deepens, the diagonalization curve approaches the cosine fit, ie.~approaches the tight-binding limit as we'd expect. In the less extreme
case (Fig.~\ref{fig:tightbind}b) an asymmetry exists, and electron effective masses (related to the band curvature at $K_y=0$) are larger than
hole effective masses (related to the band curvature at $K_y=\pi/a$). For a deeper investigation of this electron-hole asymmetry topic, see Section IV.D of Ref.~[\onlinecite{pavelich15}].

\section{MUFFIN-TIN POTENTIAL}
\label{sec:muffin}

We have used square potentials thus far; however, more realistic potentials arise through central forces. Therefore a more realistic
potential is the so-called ``muffin-tin potential," named for its resemblance to the depression in a muffin tin tray. Where the 2D Kronig-Penney potential was a repeating series of square wells in a square lattice, a muffin-tin potential is one with cylindrical wells. Such a potential imposes no conceptual difficulties, and the matrix elements are readily expressed as (as in Section \ref{sec:2dkp} we are using $a_x = a_y \equiv a$)
\begin{equation}
H_{n_xn_y,m_xm_y}^V = \frac{V_0}{a^2} \int_{\frac{a}{2}-r}^{\frac{a}{2}+r} \dif x \int_{\frac{a}{2}-\sqrt{r^2-(x-\frac{a}{2})^2}}^{\frac{a}{2}+\sqrt{r^2-(x-\frac{a}{2})^2}} \dif y\,  e^{i2\pi \left(m_x - n_x\right)x/a} e^{i2\pi \left(m_y - n_y\right)y/a}.
\label{eq:muffintinHV}
\end{equation}
The integral bounds come from the equation for a circle with radius $r$ centered at $(x,y) = \left(\frac{a}{2},\frac{a}{2}\right)$, namely
\begin{equation}
\left(x - \frac{a}{2}\right)^2 + \left(y - \frac{a}{2}\right)^2 = r^2.
\end{equation}
The well depth is $V_0$ and will typically have a negative value. Positive $V_0$ would result in a series of columns. By making dimensionless substitutions $x/a \rightarrow x$, $y/a \rightarrow y$ as before, and defining $\bar{r} \equiv r/a$, we get 
\begin{equation}
h_{n_xn_y,m_xm_y}^V = {v_0} \int_{\frac{1}{2}-\bar{r}}^{\frac{1}{2}+\bar{r}} \dif x \int_{\frac{1}{2}-\sqrt{\bar{r}^2-(x-\frac{1}{2})^2}}^{\frac{1}{2}+\sqrt{\bar{r}^2-(x-\frac{1}{2})^2}} \dif y\,  e^{i2\pi \left(m_x - n_x\right)x} \, e^{i2\pi \left(m_y - n_y\right)y}.
\end{equation}
While the inner $y$ integral is easily evaluated, we know of no analytic solution
for the general matrix element. For the main diagonal matrix elements the exponentials reduce to unity and so the contribution is just the volume of a cylinder $\pi \bar{r}^2 v_0$. For the off-diagonal elements we simply compute them numerically.\cite{matlabintegral2}

At this point the reader may ask: if the matrix elements are so readily computable numerically, why do we bother finding analytic forms for the matrix elements? A few points are worth considering:
\begin{itemize}
\item \textit{Computational efficiency}. Analytic expressions can typically be evaluated much faster than numerical integrations, since our particular implementation involves rapidly oscillating basis states for larger matrices.
\item \textit{Mathematical insight}. Numerical results are black boxes, whereas analytic expressions may suggest new interpretations of the problem.
\item \textit{Numerical error}. Analytic expressions can avoid errors introduced from numerical procedures; for example, integrating over sharply changing features like walls or cusps.
\item \textit{Pedagogy}. The calculation of the analytic expressions can be a useful exercise in understanding the matrix mechanics methodology.
\end{itemize}

\begin{figure}[h]
\centering
\subfloat[$v_0 = 0$]{\includegraphics[width=0.45\textwidth]{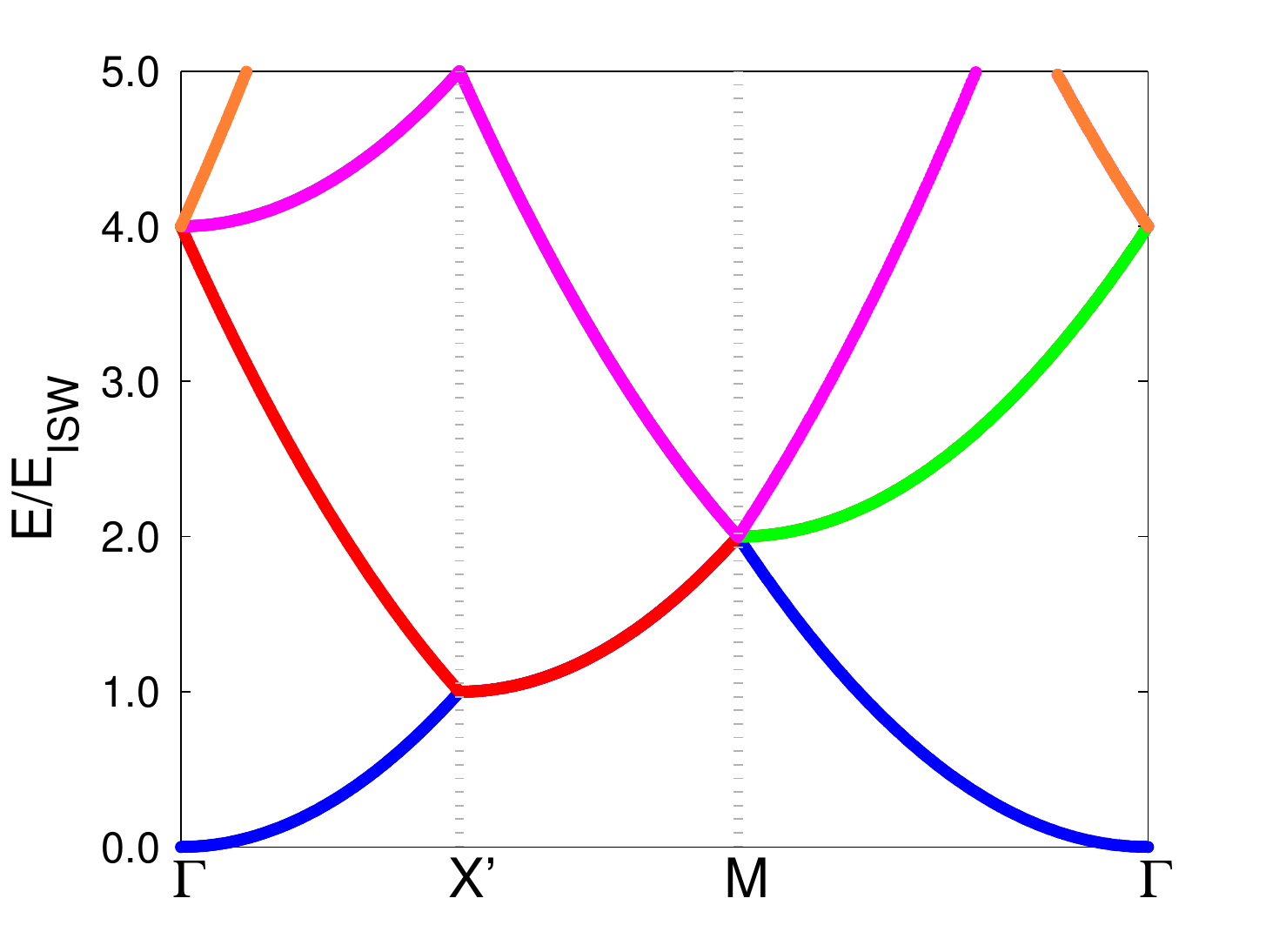}}\\
\subfloat[$v_0 = -1$]{\includegraphics[width=0.45\textwidth]{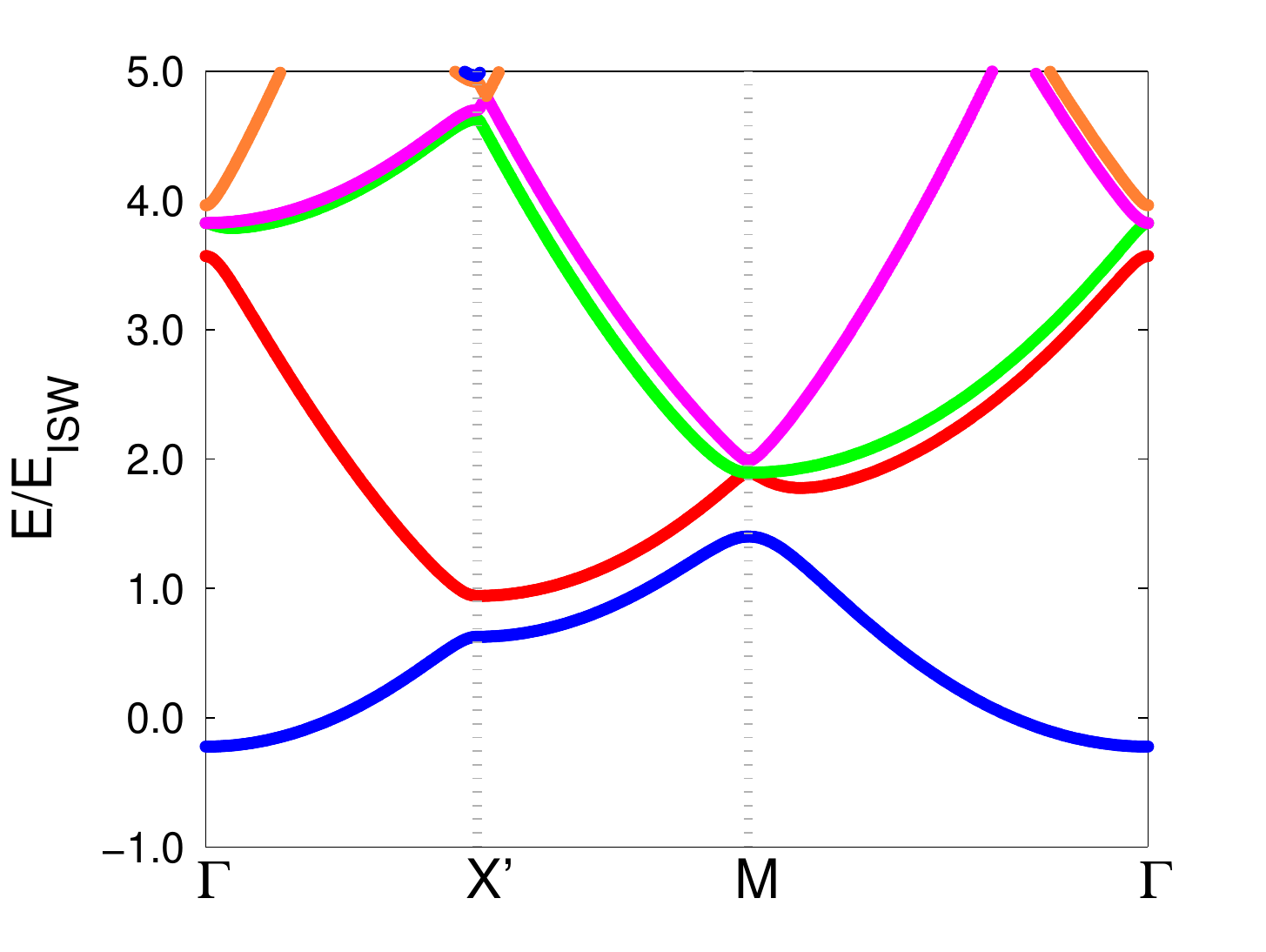}}\\
\subfloat[$v_0 = -3$]{\includegraphics[width=0.45\textwidth]{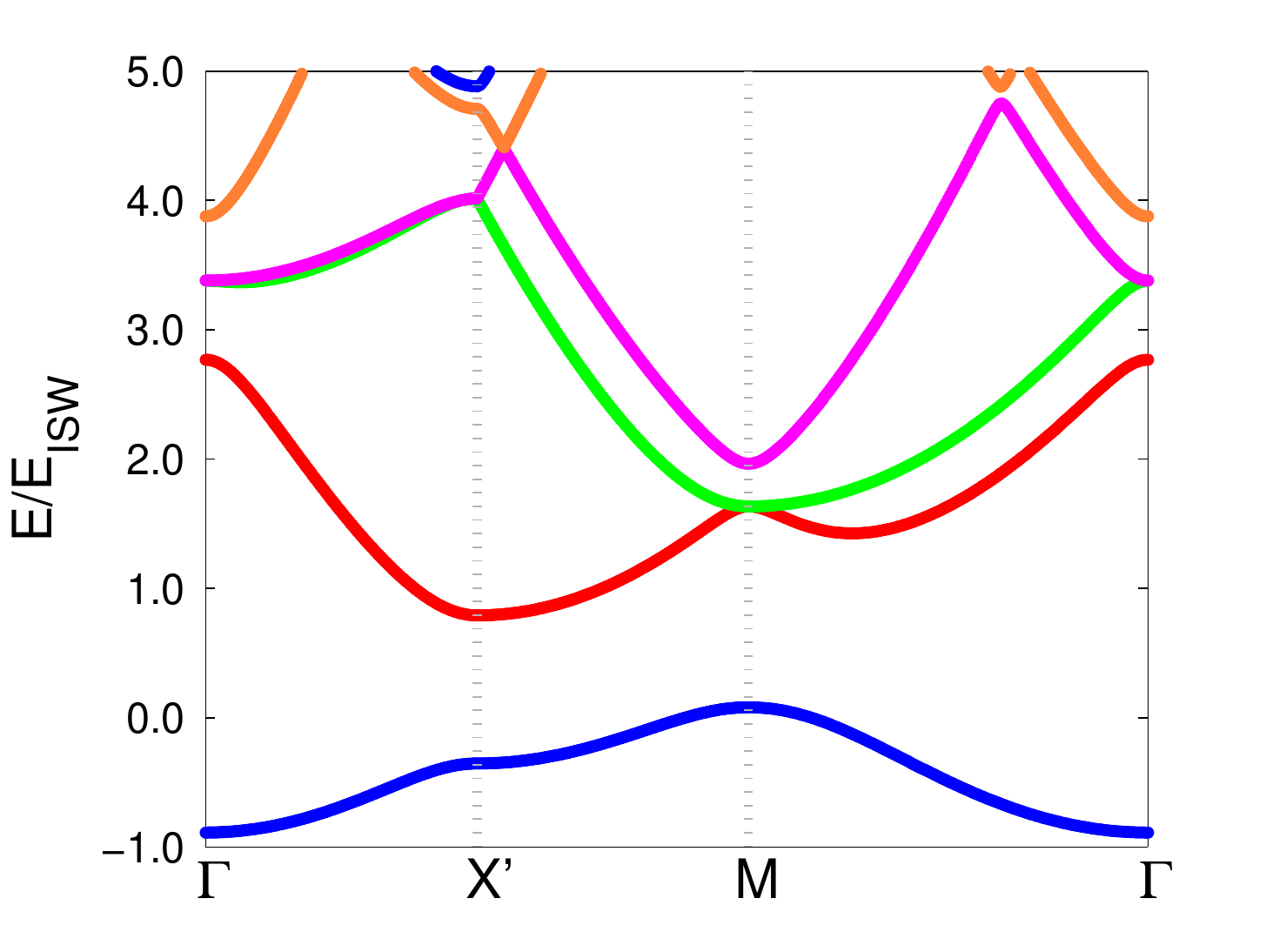}}
\caption{Generated band structures for various muffin-tin well depths with $\bar{r} = 0.25$.}
\label{fig:muffintin}
\end{figure}

Some band structures for the muffin-tin potential are shown in Fig.~\ref{fig:muffintin}. These results compare favorably to other methods found in the literature (see, for example, Fig. 5 in Ref.~[\onlinecite{Liu2003}]). It takes appreciably more time to generate these figures than the 2D Kronig-Penney figures.

\section{GAUSSIAN POTENTIAL}
\label{sec:gauss}

The potentials mentioned so far have vertical walls with constant well depths. A more realistic potential would be smoothly varying, and so we next turn to the two-dimensional Gaussian potential which has the form (again $a_x = a_y \equiv a$)
\begin{equation}
V(x,y) = V_0 \, \exp\left[{ -\left( \alpha_x\frac{(x-x_0)^2}{a^2} + \alpha_y\frac{(y-y_0)^2}{a^2}\right)}\right]
\end{equation}
where $V_0$ represents the maximum depth of the well (and therefore typically takes on a negative value), $(x_0,y_0)$ are the coordinates of the center of the well, and $\alpha_x$ and $\alpha_y$ are measures of the ``range" of the well in either direction.

Like the 2D Kronig-Penney potential and unlike the muffin-tin potential, the Gaussian readily factorizes into two separate one-dimensional integrals. First we cast the problem into dimensionless form as we did in sections \ref{sec:2dkp} and \ref{sec:muffin} and then we can write the matrix elements just like in Eq.~\ref{eq:2dkronigfactored} as
\begin{equation}
h_{n_xn_y,m_xm_y}^V = v_0 I(n_x, m_x, \alpha_x) I(n_y, m_y, \alpha_y)
\end{equation}
where
\begin{equation}
I(n,m,\alpha) = \int_0^1 \dif x \, e^{-\alpha \left( x-x_0 \right)^2} e^{i2\pi (m-n) x}.
\end{equation}
This integral can be solved to give
\begin{equation}
I(n,m,\alpha) = \frac{1}{2} \sqrt{\frac{\pi}{\alpha}} e^{-\pi (m-n) \left[ \pi(m-n) - 2ix_0\alpha\right]/\alpha} \left( \text{erf}\left[ \frac{i\pi (m-n)  + x_0\alpha}{\sqrt{\alpha}}\right] - \text{erf}\left[ \frac{i\pi (m-n) + (x_0-1)\alpha}{\sqrt{\alpha}}\right]\right)
\label{eq:integralwitherrors}
\end{equation}
where
\begin{equation}
\text{erf}(z) = \frac{2}{\sqrt{\pi}} \int_0^z \dif t \, e^{-t^2}
\end{equation}
is the error function. Though not strictly in a reduced analytical form, mathematics packages like {Mathematica} and {MATLAB} include built-in error function routines.\cite{mathematicaerf,matlaberf} For sufficiently large $\alpha$, the error function terms in parentheses in Eq.~\ref{eq:integralwitherrors} can be approximated as $\text{erf}(\infty) - \text{erf}(-\infty) = 2$. Otherwise, care should be taken to avoid overflowing the error function.

\begin{figure}[h]
\centering
\subfloat[$v_0 = 0$, $\alpha = 1$]{\includegraphics[width=0.45\textwidth]{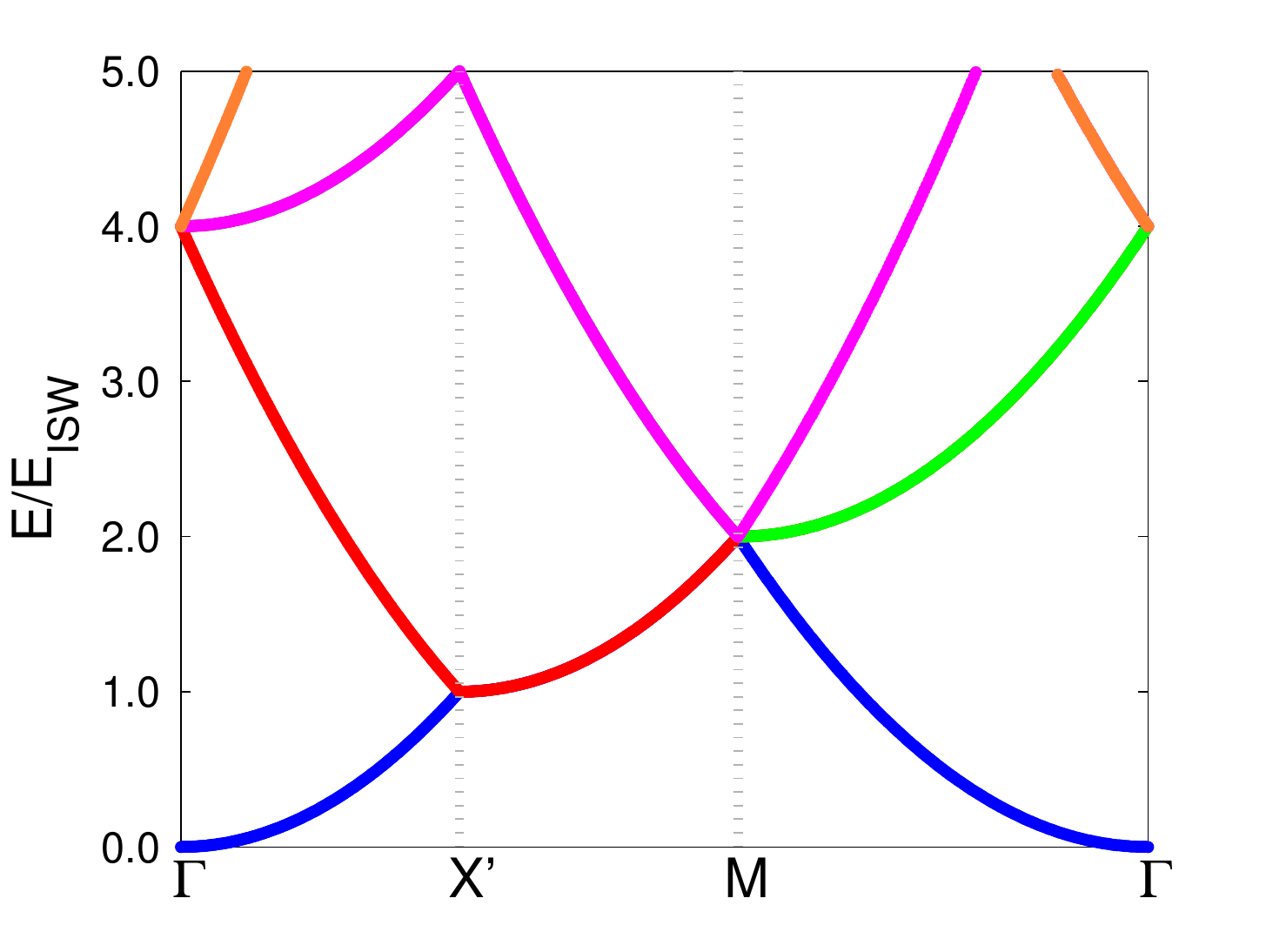}} 
\subfloat[$v_0 = 0$, $\alpha = 5$]{\includegraphics[width=0.45\textwidth]{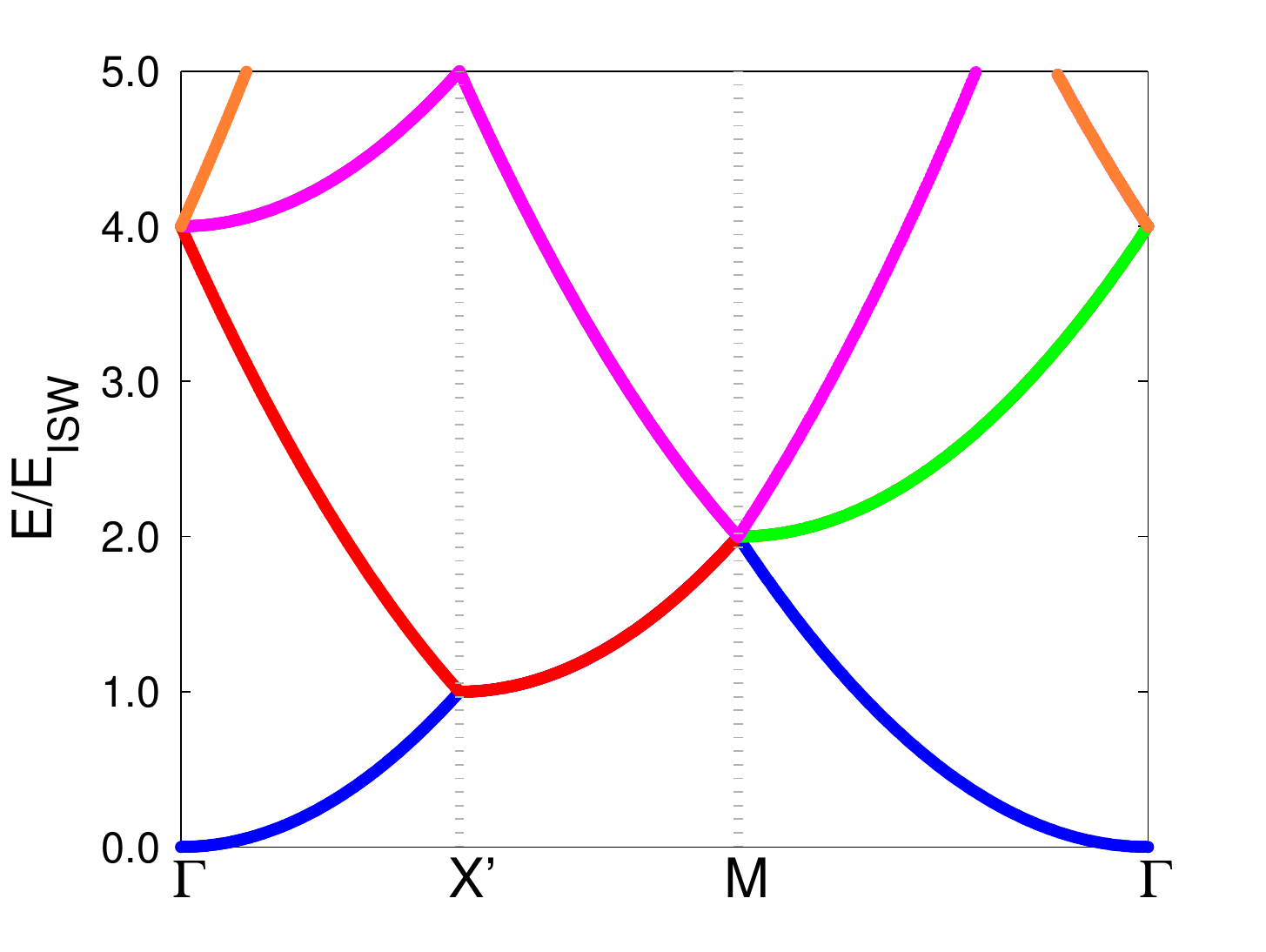}}\\
\subfloat[$v_0 = -3$, $\alpha = 1$]{\includegraphics[width=0.45\textwidth]{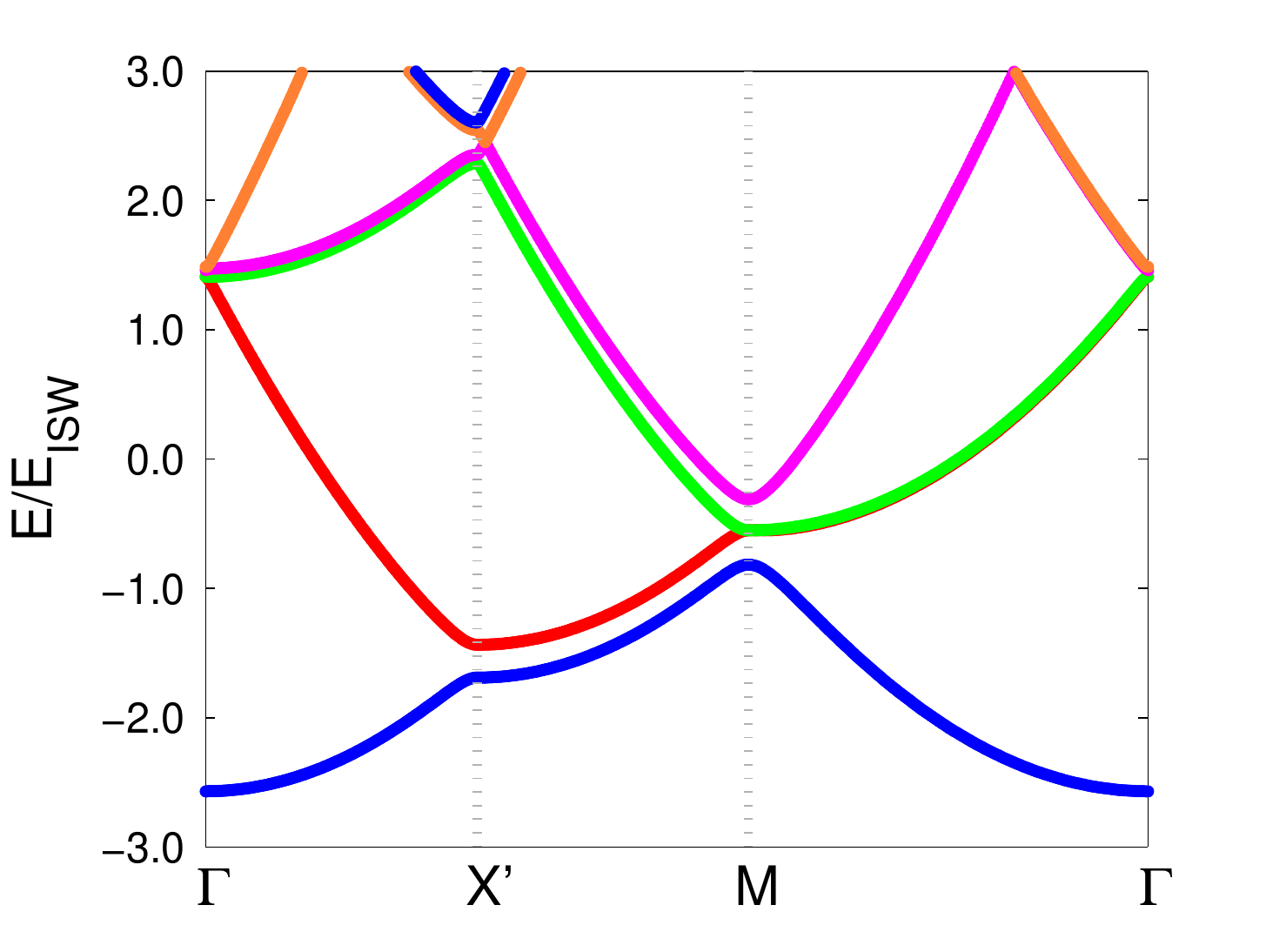}} 
\subfloat[$v_0 = -3$, $\alpha = 5$]{\includegraphics[width=0.45\textwidth]{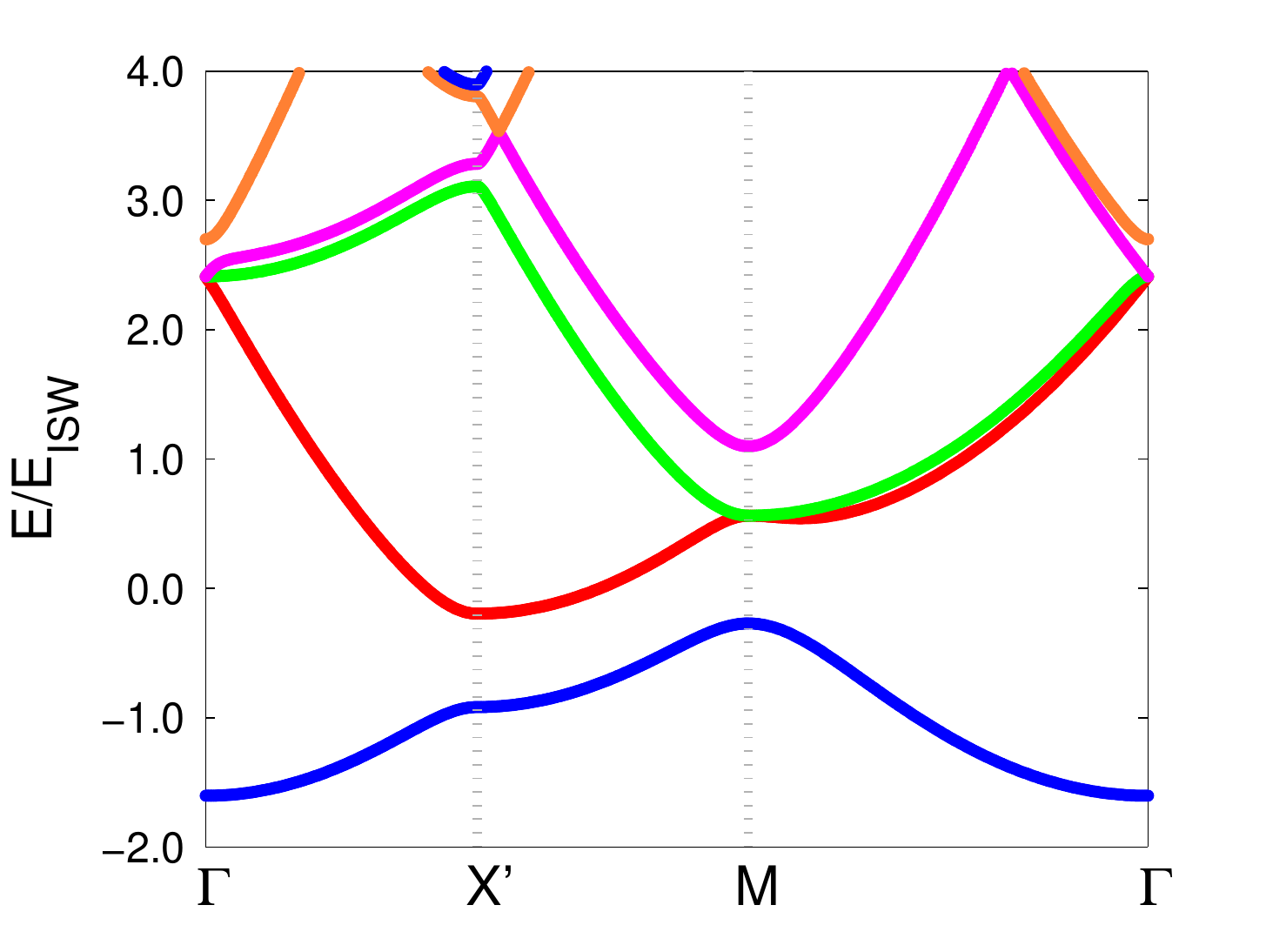}}\\
\subfloat[$v_0 = -10$, $\alpha = 1$]{\includegraphics[width=0.45\textwidth]{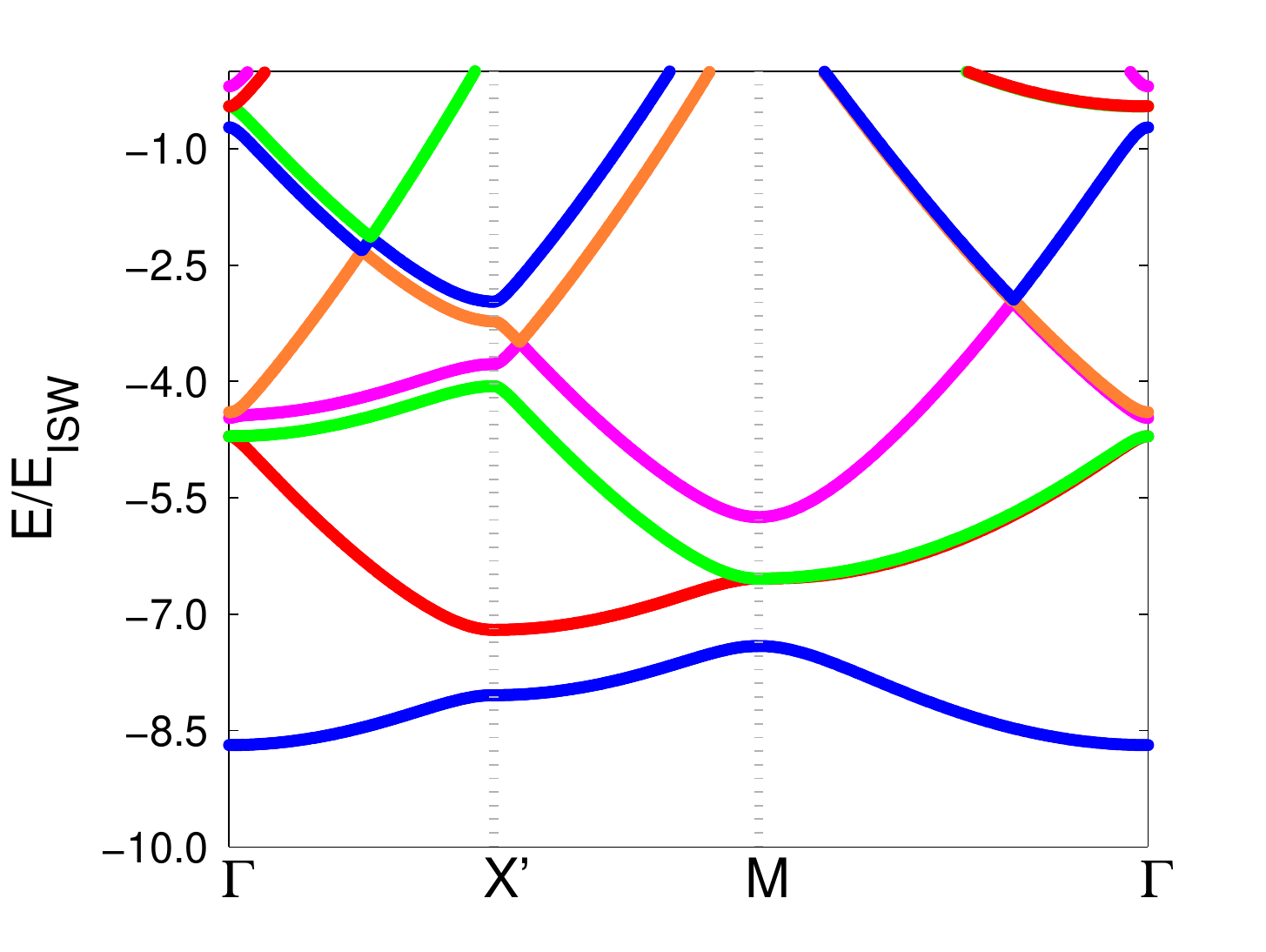}}
\subfloat[$v_0 = -10$, $\alpha = 5$]{\includegraphics[width=0.45\textwidth]{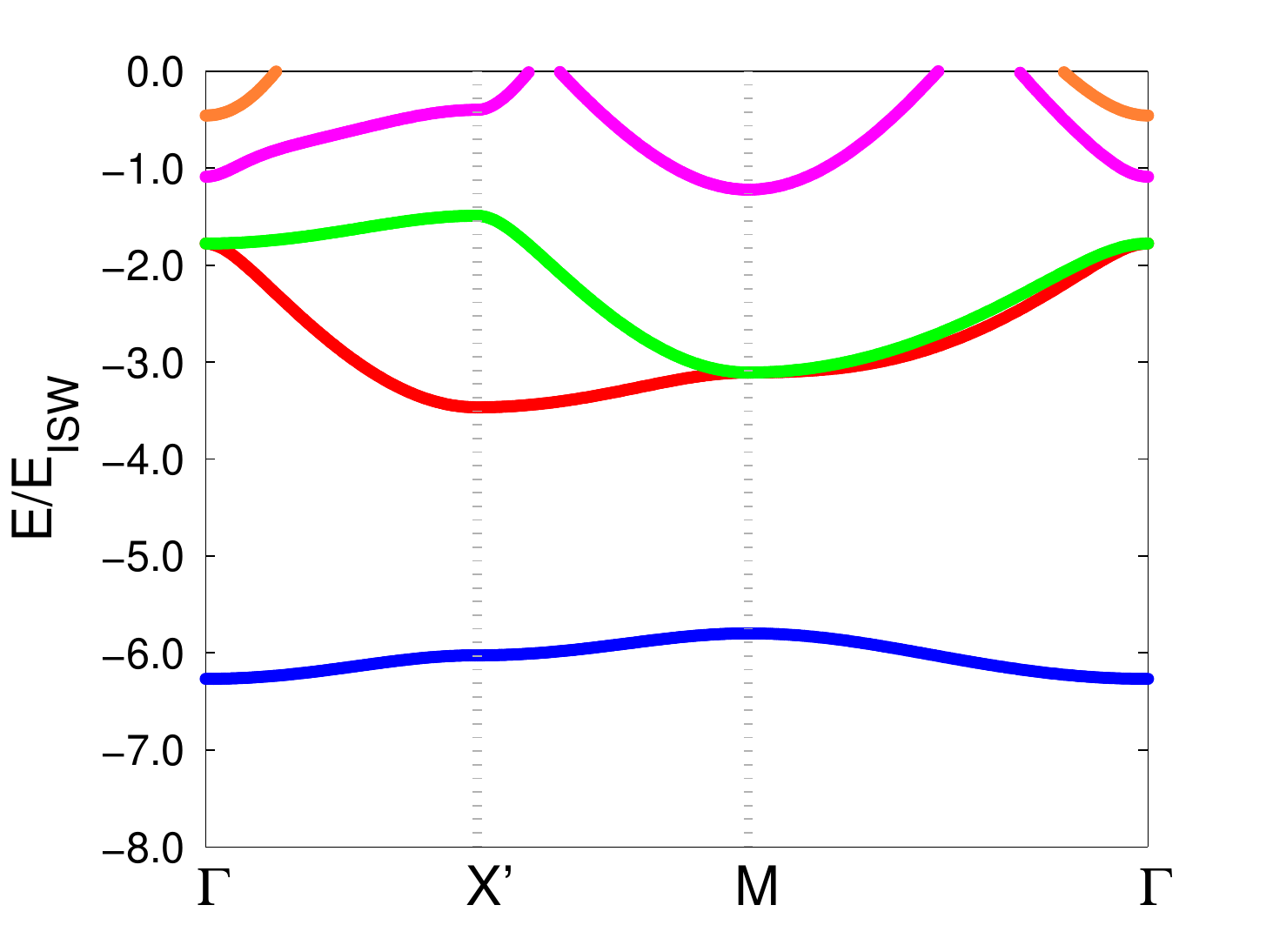}}
\caption{Generated band structures for various Gaussian wells. Here $\alpha_x = \alpha_y = \alpha$ and $(\frac{x_0}{a},\frac{y_0}{a}) = (\frac{1}{2}, \frac{1}{2})$. There are no significant differences with the bands generated with muffin-tin potentials in Fig.~\ref{fig:muffintin}.}
\label{fig:gaussbands}
\end{figure}

Plots of the band structure for various values of $v_0$ and $\alpha$ (where we have set $\alpha_x = \alpha_y = \alpha$ and placed the center of the well at the center of the unit cell) are shown in Fig.~\ref{fig:gaussbands}. There are no significant qualitative differences in the band structure obtained with this potential vs. the muffin tin potential, but
comparison of Fig.~\ref{fig:muffintin} with Fig.~\ref{fig:gaussbands} shows clear quantitative differences.

While we have shown how easily the formalism can handle a case like Gaussian wells, we are in general hesitant to use such potentials that do not vanish at the boundaries of our unit cell. There will be sharp cusps in the potential at the unit cell boundaries, whereas we expect smooth wraparound in the cell since there is nothing intrinsically special about the boundaries when we obey the general periodicity condition in Eq.~\ref{eq:2dperiodicity}. Thus, our mathematical model may not adhere to our desired smoothly-varying physical model.

It is for this reason we also avoid the case of the 2D pseudo-Coulomb potential
\begin{equation}
V(x,y) = \frac{-V_0}{\sqrt{\left( x - a/2\right)^2 + \left( y - a/2\right)^2 + b^2}}
\end{equation}
where $b$ is some parameter introduced to prevent a singularity at the center. There are no further difficulties with the methodology for this model, however. 

\section{TWO-ATOM UNIT CELL}
\label{sec:twocell}

We have thus far limited ourselves to one atom per unit cell, but it is of interest for more complicated structures that we relax this restriction. To begin, we consider again the 2D Kronig-Penney model but viewed through a larger ``window," with a rectangle consisting of two of the previously discussed unit cells. That is, we'll consider a rectangle with side lengths $a_x$ and $a_y$ where $a_x = 2a_y$ for the $x$- and $y$-directions respectively, with two square wells centered at $(x_1, y_1) = (a_x/2, a_y/2)$ and $(x_2, y_2) = (3a_x/2, a_y/2)$. Physically,
this is identical to the model with a square unit cell considered in section III.

We can then compute for the Hamiltonian elements
\begin{equation}
H_{n_xn_y,m_xm_y}^V = \frac{V_0}{a_xa_y} \left(\int_{a_x/4}^{3a_x/4}  + \int_{5a_x/4}^{7a_x/4} \right) \dif x \, e^{i2\pi \left(m_x - n_x\right)x/a_x} \int_{a_y/4}^{3a_y/4} \dif y \, e^{i2\pi \left(m_y - n_y\right)y/a_y}
\end{equation}
or in dimensionless form
\begin{eqnarray}
h_{n_xn_y,m_xm_y}^V &=& v_0 \left(\int_{1/4}^{3/4} + \int_{5/4}^{7/4} \right) \dif x \, e^{i2\pi \left(m_x - n_x\right)x} \int_{1/4}^{3/4} \dif y \, e^{i2\pi \left(m_y - n_y\right)y}.
\end{eqnarray}
We have written the $x$ integrals in a slightly peculiar way; all we mean to say is that we compute two integrals with the separate bounds of the two wells but with the same integrand, and the $y$ integral distributes over these two multiplicatively.

We now rewrite Eq.~\ref{eq:2denergyscaled} with the substitution $a_x = 2a_y$ giving
\begin{align}
E_{n_x}^{(0)} =& E_{\text{ISW}} \left( 2n_x + \frac{K_xa_x}{\pi} \right)^2 \nonumber \\
E_{n_y}^{(0)} =& E_{\text{ISW}} \left( 4n_y + \frac{K_ya_x}{\pi} \right)^2.
\label{eq:twocellenergy}
\end{align}
We have written both energy components in terms of just one length parameter (in this case, arbitrarily $a_x$) for computational convenience. We recall that while we have written Eq.~\ref{eq:twocellenergy} as two separate terms for clarity, there is only one energy level $E_{n_xn_y}^{(0)} = E_{n_x}^{(0)} + E_{n_y}^{(0)}$ for any given $(K_xa_x,K_ya_x)$.

\begin{figure}[h]
\centering
\includegraphics[scale=1.0]{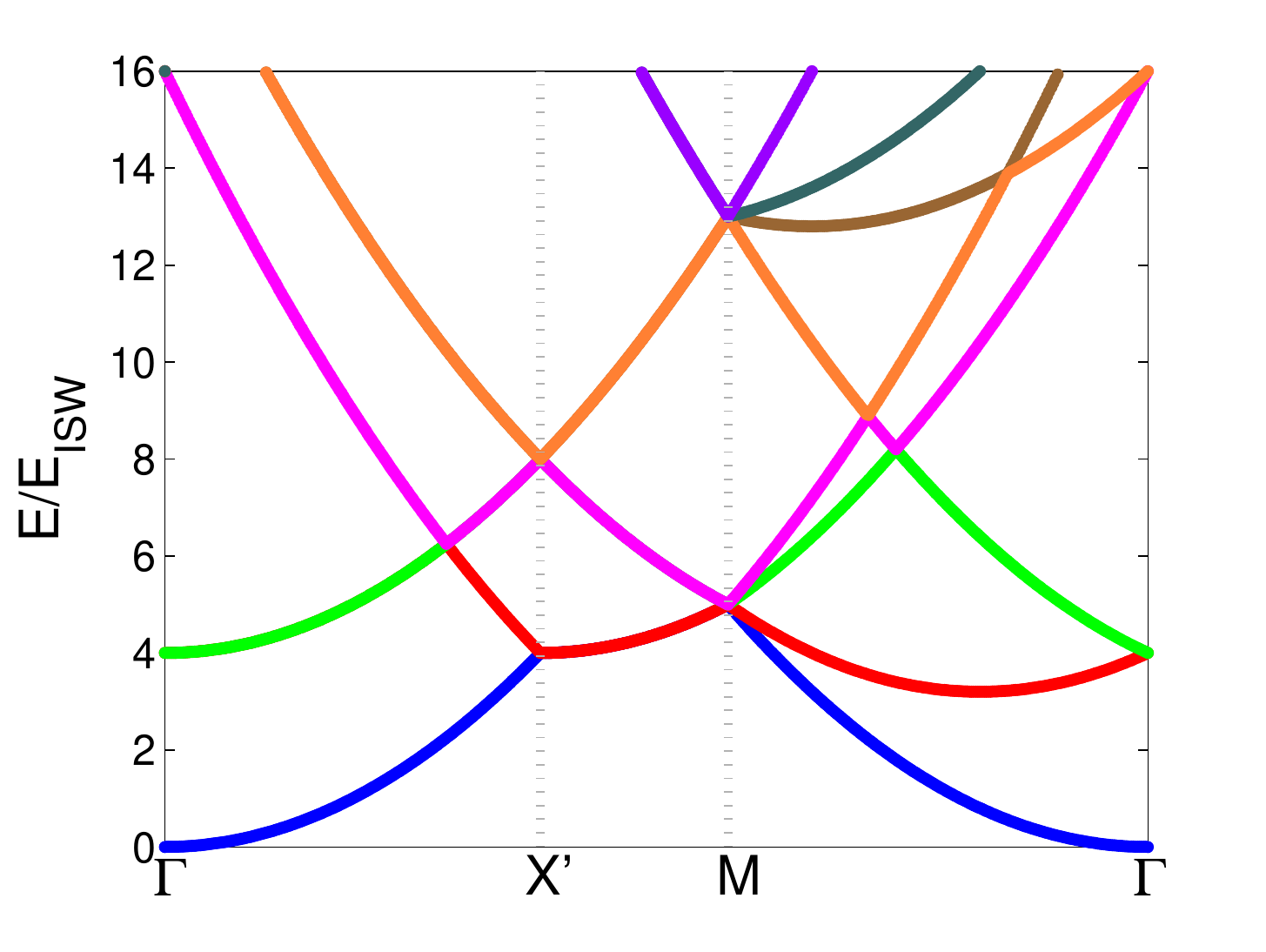}
\caption{Generated band structure for the two-site rectangular cell for $v_0 = 0$.}
\label{fig:twocellbands}
\end{figure}

In Fig.~\ref{fig:twocellbands} we show the generated band structure for $v_0 = 0$. Now, $X' = (0, \pi/a_y) = (0, 2\pi/a_x).$ As usual we have used colors to distinguish when eigenenergies are being plotted, but since the electronic branches cross each other in places we see that the eigenenergy ordering does not correspond necessarily to physical meaningful branch ordering.

Clearly, Fig.~\ref{fig:twocellbands} is different from Fig.~\ref{fig:kpbands}b. This is a general problem in electronic band structure calculations where the use of a non-primitive unit cell (i.e.~not the Wigner-Seitz cell) produces a band diagram that contains extra information compared to the first Brillouin zone corresponding to the primitive cell.\citep{boykin2009} Primitive cells, however, are often complicated geometrically with difficult-to-satisfy boundary conditions, making calculations more intractable. Techniques exist to ``zone unfold" the band diagrams from non-primitive cells into their Brillouin zone analogues.\citep{farjam2015} Applying a zone unfolding procedure is outside the scope of this paper, but in Section \ref{sec:hexagonal} we will compare our method's output to a pre-unfolded result in the literature.

\section{HEXAGONAL LATTICE}
\label{sec:hexagonal}

\begin{figure}[h]
\centering
\subfloat[Schematic representation with correct relative sizes of the unit cell structures (square wells).]{\includegraphics[width=0.42\textwidth]
{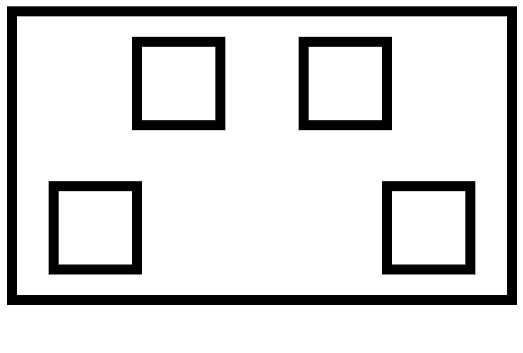} \label{fig:hexagonalcell} }
\subfloat[Countour map of the ground state wavefunction (square wells). Note the lack of hexagonal symmetry in the wave functions
due to the square potentials.]{\includegraphics[width=0.51\textwidth]
{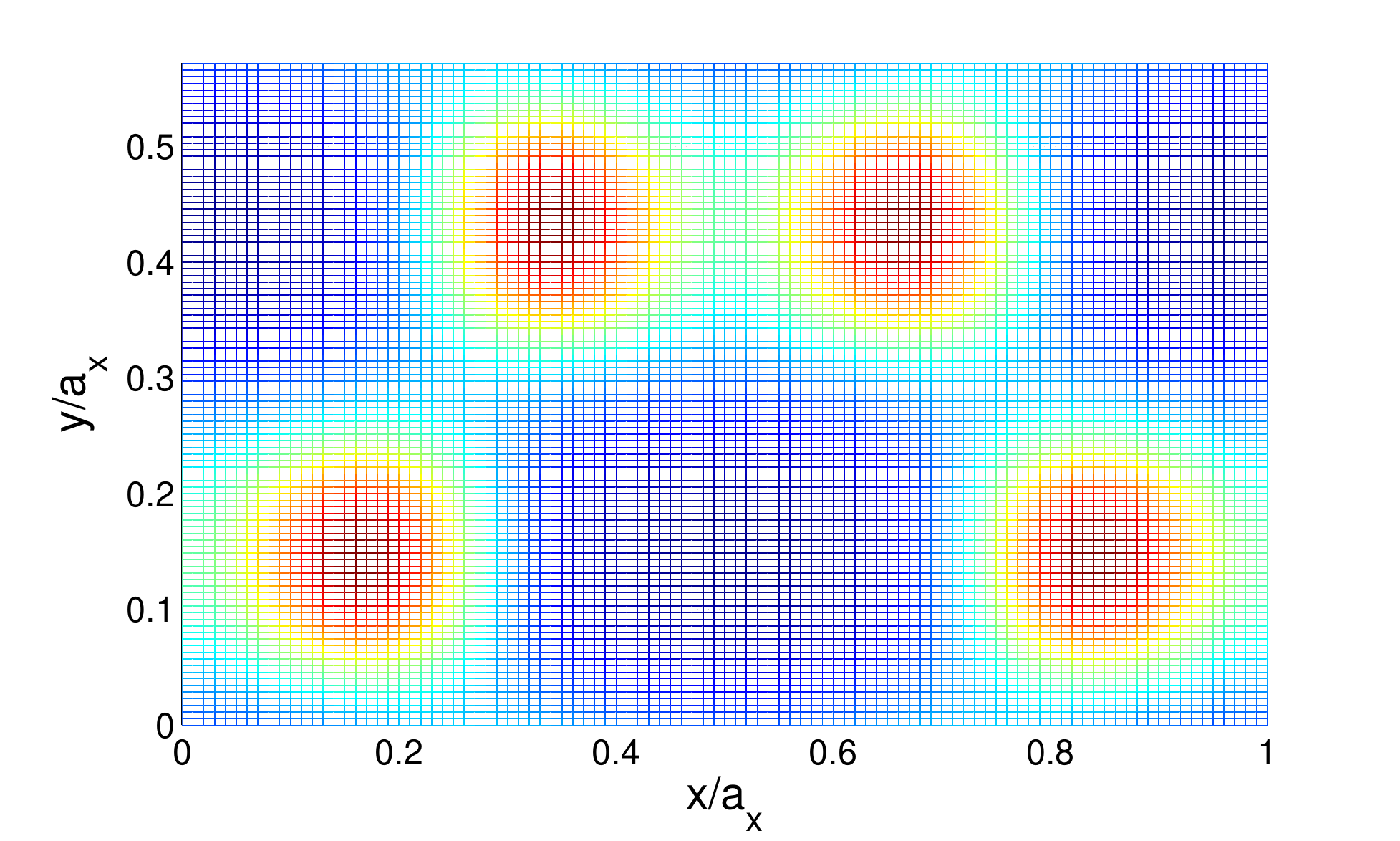} \label{fig:hexagonalwf}} \\
\subfloat[Schematic representation with correct relative sizes of the unit cell structures (muffin-tin wells).]{\includegraphics[width=0.42\textwidth]
{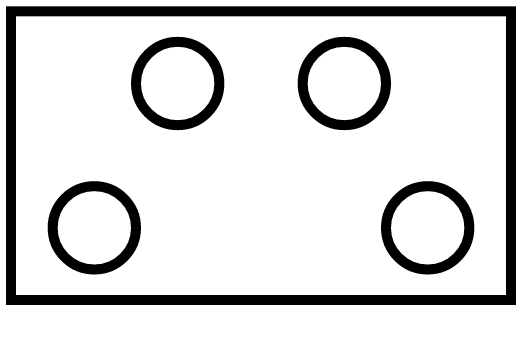} \label{fig:hexagonalcellmuffin}}
\subfloat[Countour map of the ground state wavefunction (muffin-tin wells). Hexagonal symmetry is clearly present. Compare with
(b).]{\includegraphics[width=0.51\textwidth]
{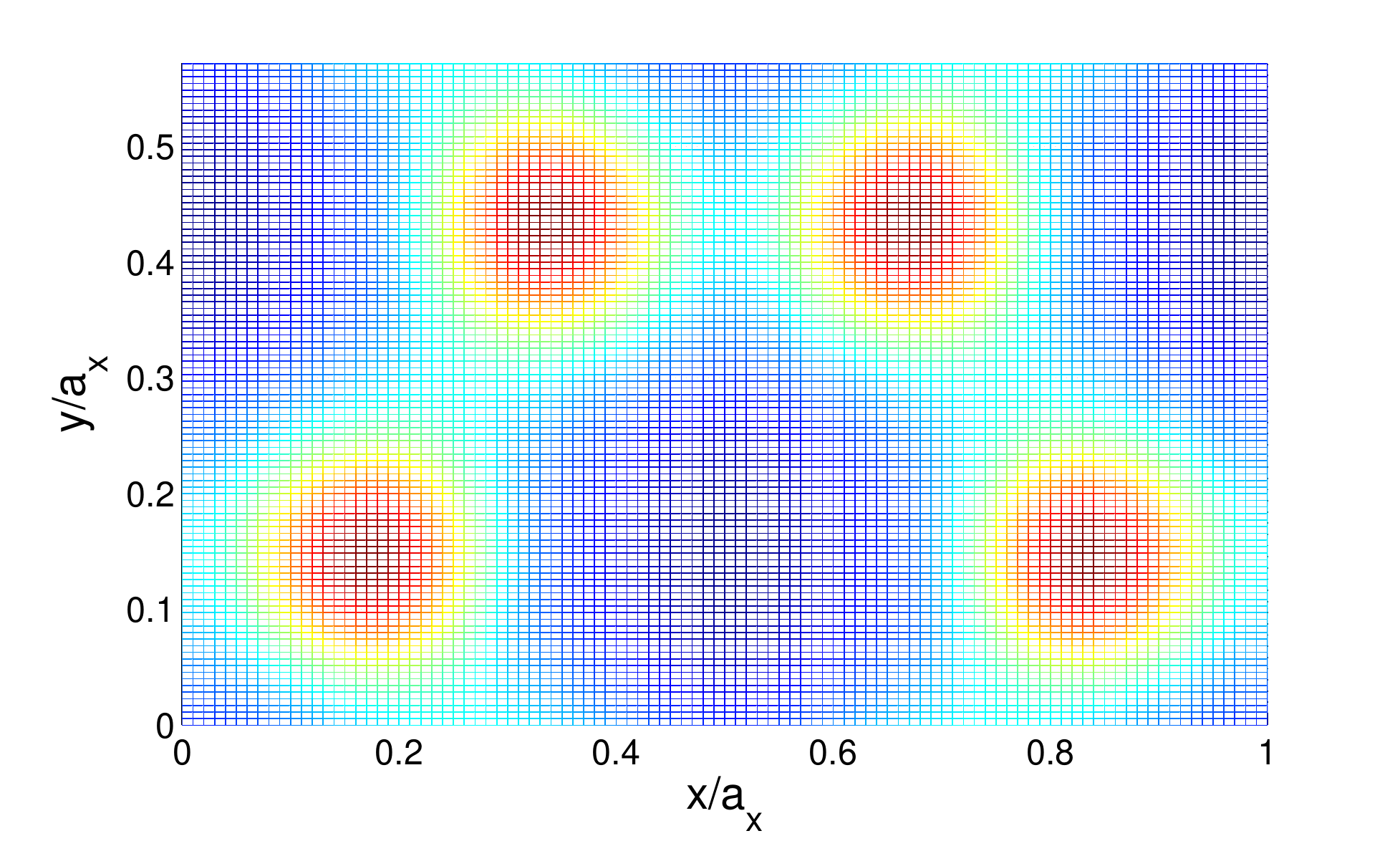} \label{fig:hexagonalwfmuffin}}
\caption{The rectangular unit cell for the hexagonal lattice, both a schematic representation of the placement of the square wells and the contour map of the ground state wavefunction. Top row is for square wells, bottom row is for muffin-tin wells. $v_0 = -5$ in both cases.}
\label{fig:hexagonal}
\end{figure}

Hexagonal lattices are typically tiled with rhomboid unit cells using a two atom-per-site basis. Our matrix mechanics method is more easily
implemented with a rectangular cell, however, in order to easily use the plane-wave basis state expansion. A rectangular tiling can be accomplished with the unit cell shown in Fig.~\ref{fig:hexagonalcell}. This is likely not immediately obvious, but laying out a grid of such cells will show the hexagonal pattern emerging. If the ``bond length" of the hexagonal lattice is called $\delta$, the dimensions of the rectangle are $3\delta \times \sqrt{3	}\delta$ with the ``atomic" sites located at $(\frac{1}{2}\delta, \frac{\sqrt{3}}{4}\delta), (\delta, \frac{3\sqrt{3}}{4}\delta), (2\delta, \frac{3\sqrt{3}}{4}\delta), (\frac{5}{2}\delta, \frac{\sqrt{3}}{4}\delta)$. Showing this is a fairly simple exercise in geometry. Our basis states will then be as in Eq.~\ref{eq:2dbasisstates} with $a_x = 3\delta$ and $a_y=\sqrt{3}\delta$. Here we have chosen square wells with a width of $\frac{1}{2}\delta$. One can then use matrix elements as per Eqs.~\ref{eq:2dkronigfactored} and \ref{eq:2dkronigI} with appropriately chosen $p$ values. Further, the relative energy scaling in Eq.~\ref{eq:2denergyscaled} must be taken into account, where we have $(a_x^2/a_y^2) = (a_x^2/(\frac{\sqrt{3}}{3}a_x)^2) = 3$, giving (when we rewrite in terms of $a_x$ as we did in Eq.~\ref{eq:twocellenergy})
\begin{align}
E_{n_x}^{(0)} =& E_{\text{ISW}} \left( 2n_x + \frac{K_xa_x}{\pi} \right)^2 \nonumber \\
E_{n_y}^{(0)} =& E_{\text{ISW}} \left( 3 \cdot 4n_y^2 + \left(\frac{3}{\sqrt{3}}\right)4 \frac{n_y K_y a_x}{\pi} + \left(\frac{K_ya_x}{\pi}\right)^2 \right) \nonumber \\
=& E_{\text{ISW}} \left( 2\sqrt{3}n_y + \frac{K_ya_x}{\pi} \right)^2.
\end{align}

\begin{figure}[h]
\centering
\subfloat[First four bands, square wells.]{\includegraphics[width=0.375\textwidth]{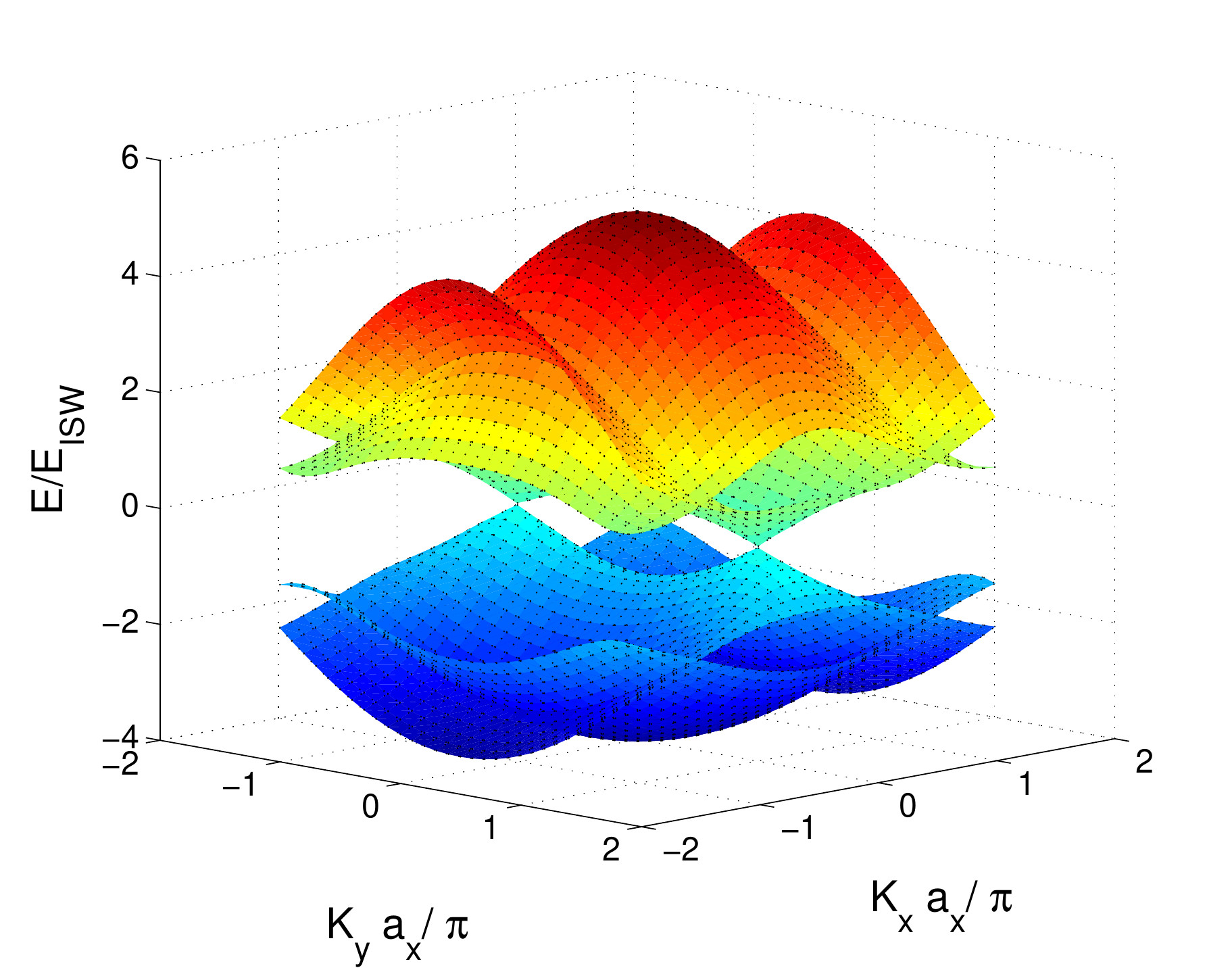} \label{fig:hexagonalbandssquare}} 
\subfloat[First four bands, muffin-tin wells.]{\includegraphics[width=0.375\textwidth]{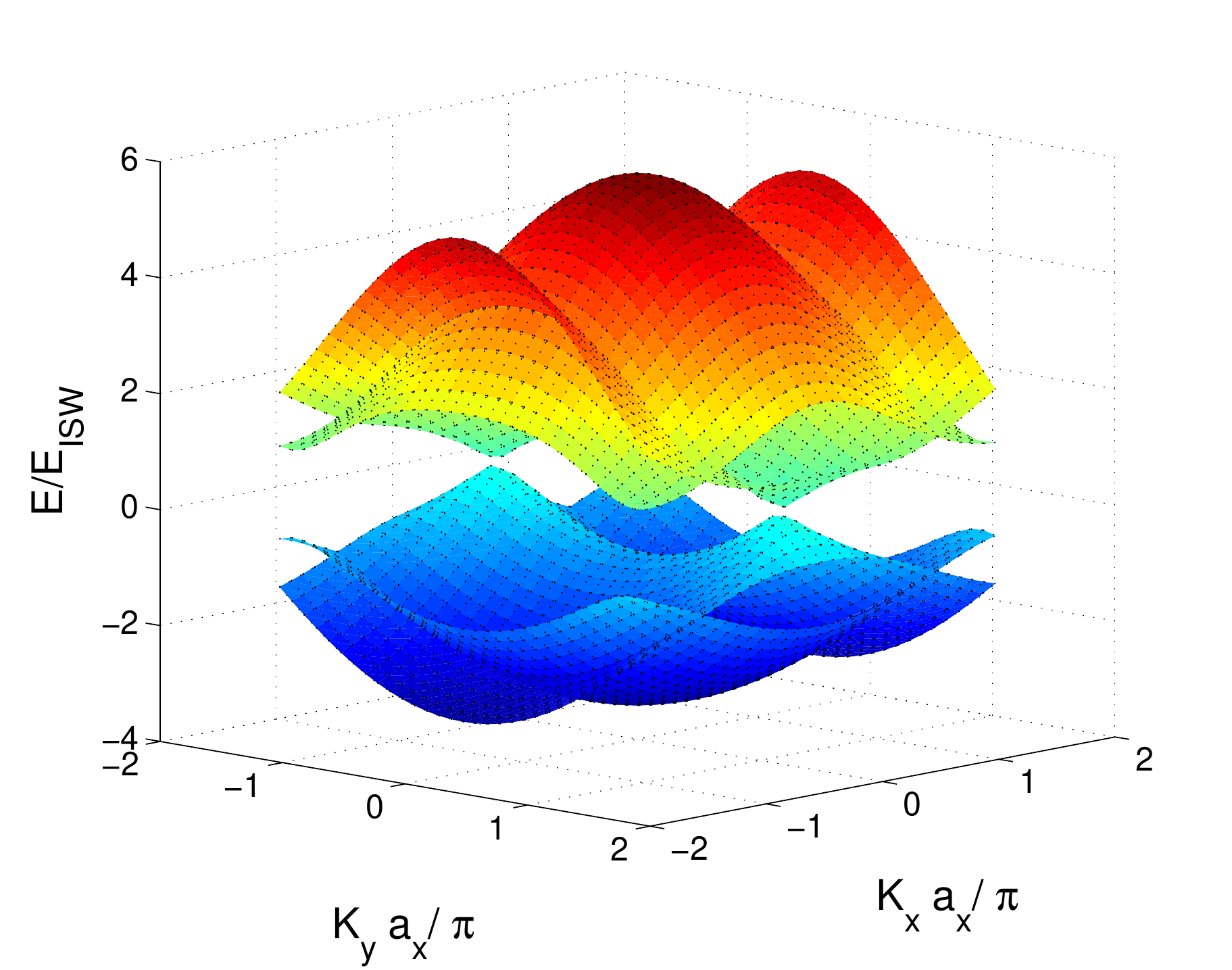} \label{fig:hexagonalbandsmuffin}}\\
\subfloat[Expansion of the second and third bands around possible Dirac cone, square wells.]{\includegraphics[width=0.375\textwidth]{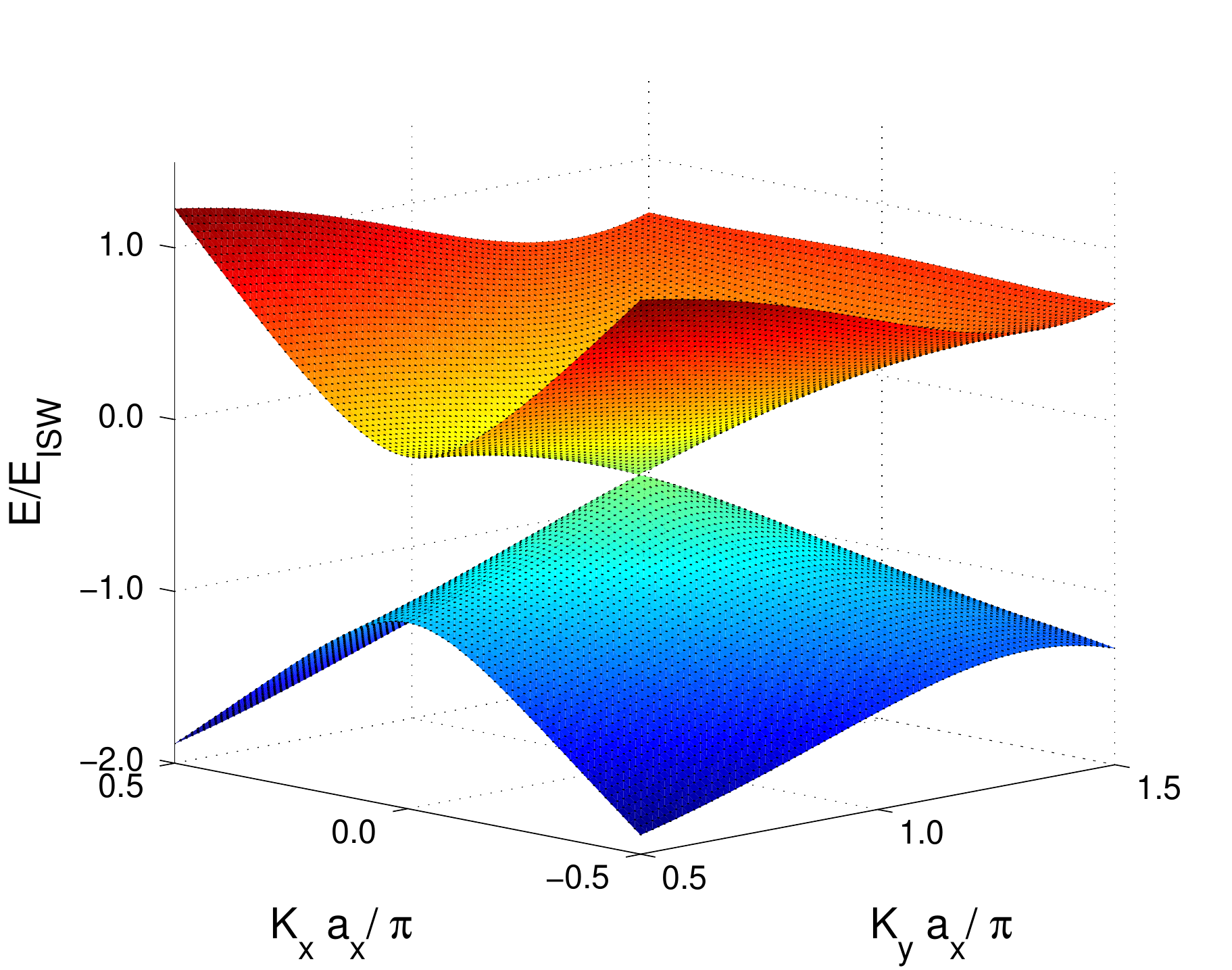} \label{fig:diracsquare}} 
\subfloat[Expansion of the second and third bands around possible Dirac cone, muffin-tin wells.]{\includegraphics[width=0.375\textwidth]{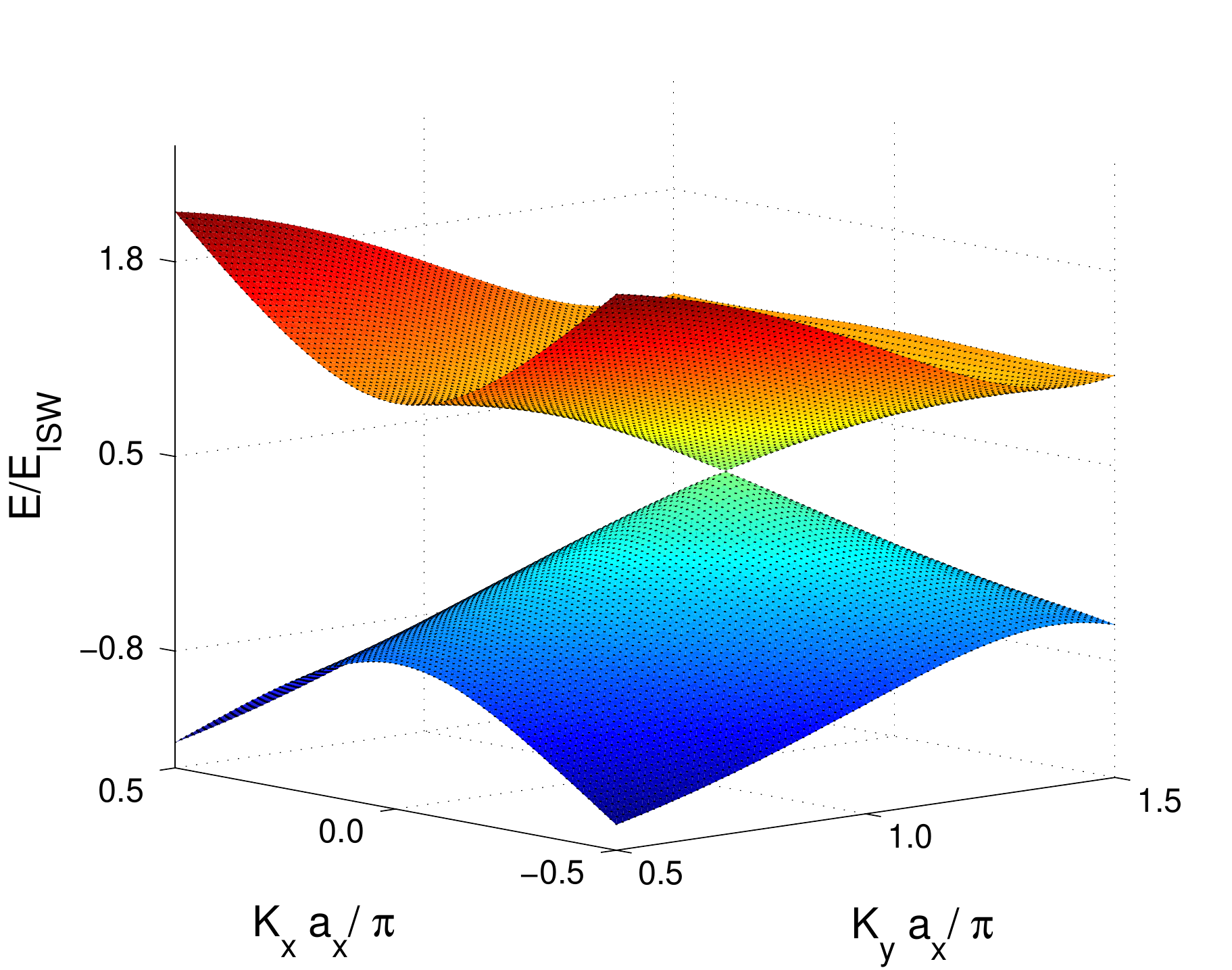} \label{fig:diracmuffin}}\\
\subfloat[Cross-section for $K_x = 0$ of the second and third bands around Dirac cone point, square wells.]{\includegraphics[width=0.375\textwidth]{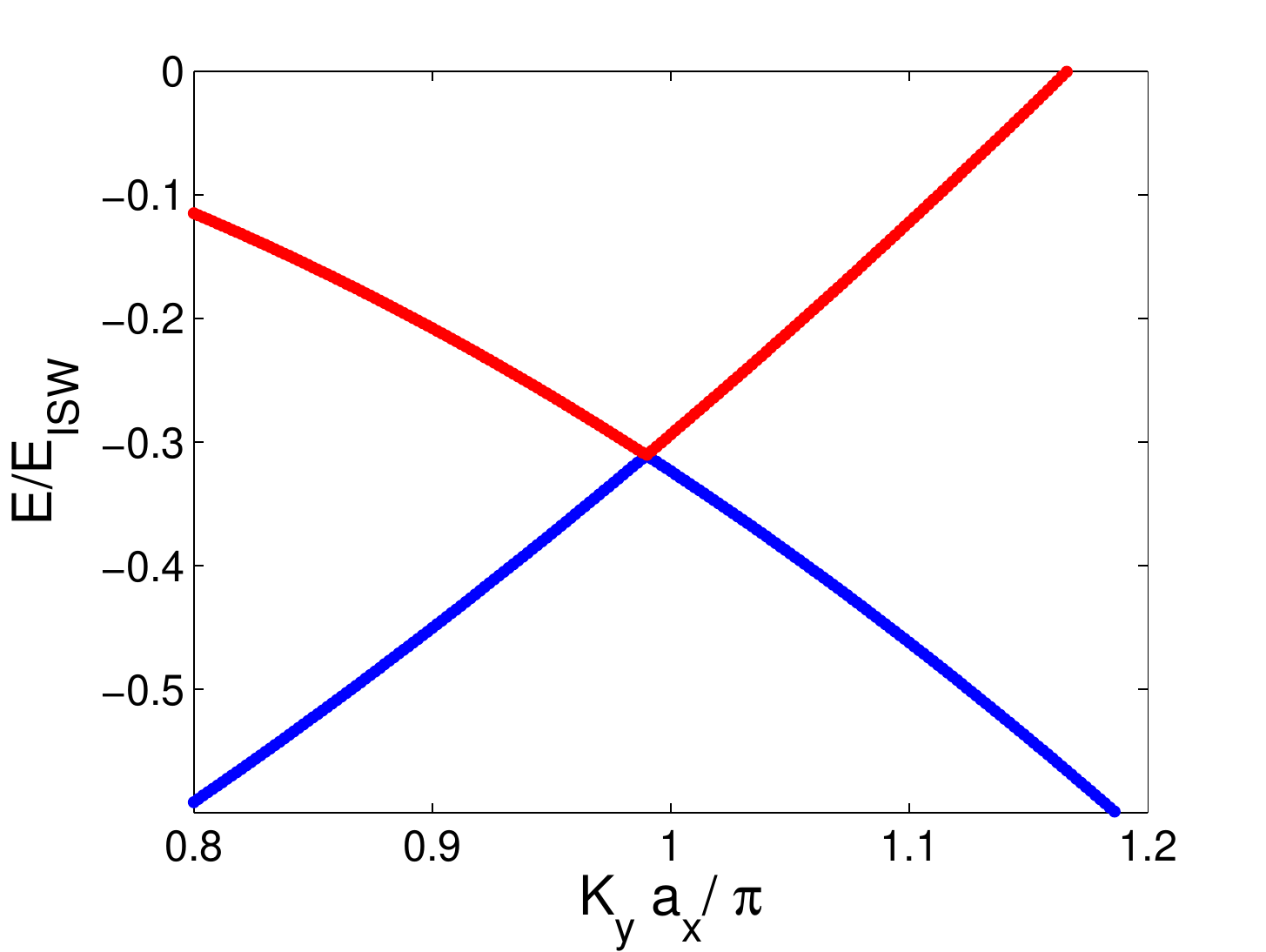} \label{fig:diracxsectionsquare}}
\subfloat[Cross-section for $K_x = 0$ of the second and third bands around Dirac cone point, muffin-tin wells.]{\includegraphics[width=0.375\textwidth]{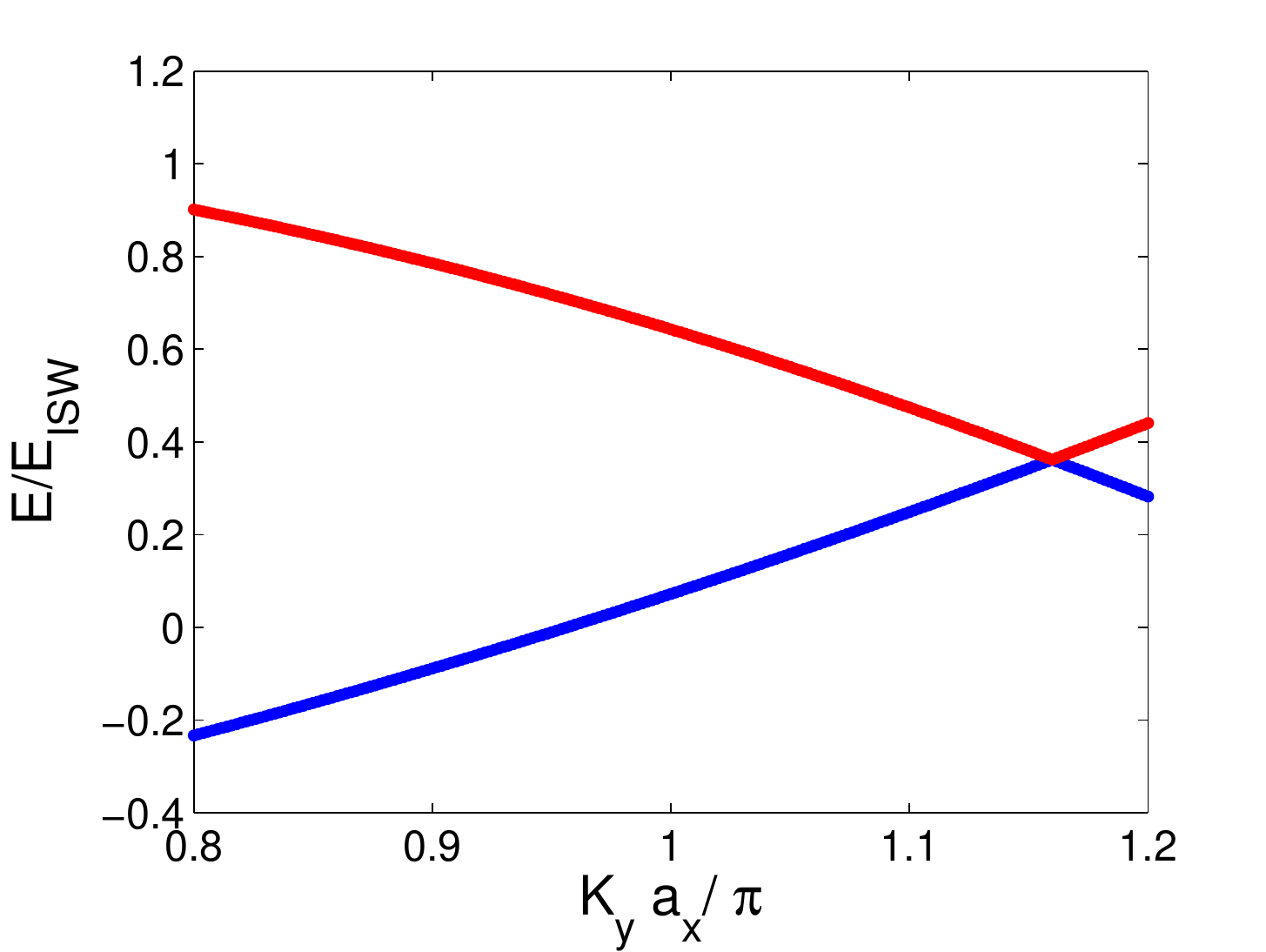} \label{fig:diracxsectionmuffin}}
\caption{Generated band structures for the hexagonal lattice, with square wells in the left column and cylindrical muffin-tin wells in the right column, both with $v_0 = -20$.}
\label{fig:hexagonalbands}
\end{figure}

From a diagonalization of such a matrix, we can reconstruct the ground state wavefunction as a check, which we have done in Fig.~\ref{fig:hexagonalwf}, showing appropriate localization in the wells. We can then use all the usual machinery to generate the energy band structure of this potential, which we have done for $v_0 = -20$ in Fig.~\ref{fig:hexagonalbandssquare}.

We see a pattern highly suggestive of Dirac cones\cite{castroneto2009} in the band structure. This is remarkable because our model is fairly simple, using only square wells. Such wells break the expected hexagonal symmetry, so a more realistic potential would have radially-symmetric wells, like the cylindrical muffin-tin wells we explored in Section \ref{sec:muffin}. This is easily done using numerical integration, though the time for computation is much slower than for the analytic square well case. 

To demonstrate the procedure, we now derive the integral for the well located at $(\frac{1}{2}\delta, \frac{\sqrt{3}}{4}\delta) = (\frac{a_x}{6}, \frac{a_y}{4}) = (\frac{a_x}{6}, \frac{\sqrt{3}a_x}{12})$. We will use $\frac{r}{a_x} = \frac{1}{12}$ to keep rough parity with the square wells. Fig.~\ref{fig:hexagonalcellmuffin} shows what such a unit cell looks like, and Fig.~\ref{fig:hexagonalwfmuffin} serves as a check where we have reconstructed the ground state wavefunction. Note that with the radially-symmetric wells we do get the expected hexagonal symmetry, as there is now no preferred direction for the wavefunction adjoining the wells. 

The equation for such a circle is
\begin{equation}
\left(x - \frac{a_x}{6}\right)^2 + \left(y - \frac{a_x\sqrt{3}}{12}\right)^2 = r^2.
\end{equation}
Like Eq.~\ref{eq:muffintinHV} we can write
\begin{equation}
H_{n_xn_y,m_xm_y}^V = \frac{V_0}{a_xa_y} \int_{\frac{a_x}{6}-r}^{\frac{a_x}{6}+r} \dif x \int_{\frac{a_x\sqrt{3}}{12}-\sqrt{r^2-(x-\frac{a_x}{6})^2}}^{\frac{a_x\sqrt{3}}{12}+\sqrt{r^2-(x-\frac{a_x}{6})^2}} \dif y\,  e^{i2\pi \left(m_x - n_x\right)x/a_x} e^{i2\pi \left(m_y - n_y\right)y/a_y}
\end{equation}
or writing $a_y = a_x\sqrt{3}/3$ and making making the equation dimensionless as we've done previously ($x/a_x \rightarrow x$, $y/a_x \rightarrow y$, and $\bar{r} \equiv r/a_x$) we get
\begin{equation}
h_{n_xn_y,m_xm_y}^V = v_0\sqrt{3} \int_{\frac{1}{6}-\bar{r}}^{\frac{1}{6}+\bar{r}} \dif x \int_{\frac{\sqrt{3}}{12}-\sqrt{\bar{r}^2-(x-\frac{1}{6})^2}}^{\frac{\sqrt{3}}{12}+\sqrt{\bar{r}^2-(x-\frac{1}{6})^2}} \dif y\,  e^{i2\pi \left(m_x - n_x\right)x} e^{i2\pi \left(m_y - n_y\right)y/(\sqrt{3}/3)}
\end{equation}

We generate band structures as shown in Fig.~\ref{fig:hexagonalbandsmuffin}. Again we see structures highly suggestive of Dirac cones,
and the results using muffin tins are qualitatively similar to the results using square wells.

We can then easily expand around the Dirac points to get a more detailed view, as shown in Figs.~\ref{fig:diracsquare} and \ref{fig:diracmuffin}. To verify that these cones have linear dispersion, we expand yet further along a cross-section through $K_x = 0$, shown in Figs.~\ref{fig:diracxsectionsquare} and \ref{fig:diracxsectionmuffin}. These structures clearly exhibit a conical shape with linear dispersion. Remarkably, these structures are present even in the relatively crude model using square wells.

\begin{figure}[h]
\centering
\includegraphics[scale=0.7]{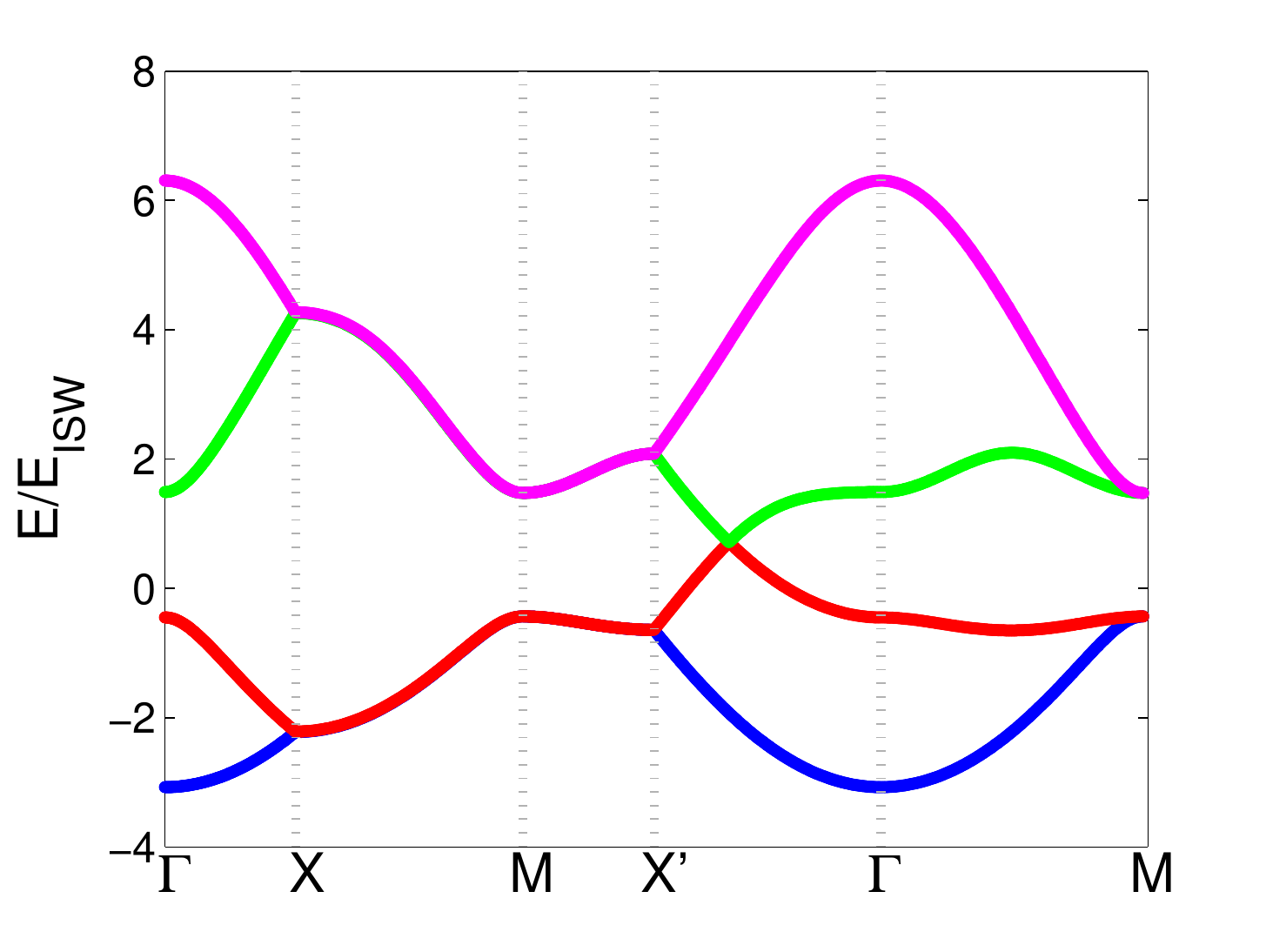}
\caption{Generated band structure for the hexagonal lattice using muffin-tin wells with $v_0 = -19$. This output is qualitatively very similar to Fig. 2d of Ref.~[\onlinecite{zheng2014}].}
\label{fig:hexagonalsymm}
\end{figure}

However, as was evident in Section \ref{sec:twocell}, our results are not directly comparable to the known locations of Dirac cones for the hexagonal lattice in the Brillouin zone. Nonetheless, we can compare our results to other work that uses a similar tiling schema for the case of a hexagonal graphene lattice. Using the rectangular high-symmetry points $\Gamma = (0,0)$, $X = (\pi/a_x,0)$, $X' = (0, \pi/a_y)$, and $M = (\pi/a_x, \pi/a_y)$ we can generate Fig.~\ref{fig:hexagonalsymm} which has excellent qualitative correspondence with Fig. 2d of Ref.~[\onlinecite{zheng2014}]. In that work such a band diagram was successfully unfolded using an unfolding program of the authors' construction. We have also obtained results for square well potentials; these are not shown as they are very similar to those shown
in Fig.~\ref{fig:hexagonalsymm}.

\section{CONCLUSION}
\label{sec:conclusion}

Previous work\cite{pavelich15} formulated a method for using matrix mechanics and plane-wave basis states to solve for the electronic band structure of one-dimensional potentials. In this paper we have extended the method to two dimensions, allowing for a much richer class of problems to be addressed. The emphasis is on conceptual clarity, where we use only relatively simple quantum mechanical techniques.
Despite the emphasis on simplicity and clarity, the method is powerful enough to investigate materials under active research, like graphene in the case of the hexagonal lattice. In future work, we plan to extend this work to three dimensions. In addition we will extend this work in two dimensions for particular lattices with interesting unit cells; or particular interest will be the breakdown of the tight-binding approximation,
prevalent in the research literature.

\section*{Acknowledgements}
One of us (FM) would like to thank Nicholas Moore for his initial work on the two-dimensional Kronig Penney model in an undergraduate
course at the University of Alberta. 
This work was supported in part by the Natural Sciences and Engineering Research Council of Canada (NSERC), by the Alberta iCiNano program, and by a University of Alberta Teaching and Learning Enhancement Fund (TLEF) grant.

\end{document}